\documentclass[a4paper,11pt,final]{article}
\usepackage[hang,flushmargin]{footmisc} % left indendt of footnote
\usepackage[titletoc,toc,title]{appendix}
\usepackage{fullpage} % changes the margin
\usepackage{xcolor}
\usepackage{titling}  % Provide double title

\makeatletter
	\renewcommand*{\@fnsymbol}[1]{\ensuremath{\ifcase#1\or\dagger\else\@ctrerr\fi}}
\makeatother

\usepackage{amsmath}
\usepackage{amssymb,wasysym}
\usepackage{dsfont}
\usepackage{upgreek}
\usepackage{mathtools}
\usepackage{enumitem}

\usepackage{siunitx}
\DeclareSIUnit\Molar{\textsc{m}}

\usepackage[version=4,arrows=pgf-filled]{mhchem}
\usepackage[outdir=./Figures/]{epstopdf}
\usepackage{graphicx}
\graphicspath{{./},{./Figures/}}
\usepackage[aboveskip=1pt,
            labelfont=bf,
            labelsep=period,
            justification=centering,
            singlelinecheck=off]{caption}
\usepackage[position=top,labelformat=simple]{subcaption}
\usepackage[export]{adjustbox}

\usepackage{natbib}
\allowdisplaybreaks

\usepackage{color}
\usepackage{xcolor}

\usepackage{listings}
\usepackage{hyperref}
\hypersetup{colorlinks,
	        linkcolor={red!50!black},
    	    citecolor={blue!50!black},
    		urlcolor={blue!80!black}}

\usepackage{siunitx}
\DeclareSIUnit\Molar{\textsc{m}}

\usepackage{ltablex}
\newcommand{\rowgroup}[1]{\hspace{1em}#1}
\let\oldhline\hline
\renewcommand{\hline}{\oldhline\noalign{\vskip .5ex}}
\newcommand\T{\rule{0pt}{2.4ex}}       % Top strut
\usepackage{floatrow}
\usepackage{lineno}
\usepackage[textsize=tiny,obeyDraft]{todonotes}
\newcommand{\HB}[1]{{\color{blue} #1}}

\newcommand{\rpar}[1]{\left(#1\right)}

\def\gs{\ensuremath{\mathbf{\upgamma}}}
\def\xn{\ensuremath{\mathbf{x}}}

\def\ip3{\ce{IP3}}

\def\pip2{\ce{PIP2}}
\def\ca{\ce{Ca^2+}}

\def\3k{\ip3\-3K}
\def\5p{IP-5P}

\def\dserine{\textsc{d}-serine}
\def\Muller{M\"{u}ller}
\def\e3{ephrin-A3}
\def\E4{EphA4}
\def\Na{\ce{Na^+}}
\def\K{\ce{K^+}}
\def\Cl{\ce{Cl^-}}
\def\H{\ce{H^+}}
\def\HB{\ce{HCO3^-}}

\def\AP{AQP4}
\def\gabaa{\ce{GABA_A}}
\def\tnfa{TNF$\upalpha$}
\def\tgfb{TGF$\upbeta$}
\def\il10{IL-10}

\newcommand*{\tabref}[1]{\tablename~\ref{#1}}
\newcommand*{\figref}[1]{\figurename~\ref{#1}}
\newcommand*{\secref}[1]{Section~\ref{#1}}

\renewcommand*{\eqref}[1]{eq.~\ref{#1}}

\newcommand*{\see}[1]{\textbf{ref.} \textsc{#1}}

\title{Neuron-Glial Interactions}
\author{
        Maurizio De Pitt\`a\\
        Basque Center for~Applied Mathematics, Bilbao, Spain\\
        mdepitta@bcamath.org
        }
\begin{document}
%%
%% Set listing style
\lstset{language=Python, basicstyle=\small\ttfamily,
              showstringspaces=false,
              columns=fixed}

%% Preliminary Title
\maketitle
\begin{minipage}{\textwidth}
Accepted for publication in the \textit{Encyclopedia of Computational Neuroscience}, D. Jaeger and R. Jung eds.,  Springer-Verlag New York, 2020 (2nd edition).
\end{minipage}

%% Title
\maketitle

\section*{Synonyms}
neuron-glial interaction modeling, synapse-astrocyte bidirectional signaling, axon-glial signaling, microglia modeling, computational glioscience

\section*{Definition}
Dependence of neuronal function on glial cells and vice verse, which consists of the interplay of potentially multiple signals that can be both of neuronal and of glial origin. This interplay may be facilitated by the specific arrangement of neurons vs. glia in different brain areas and occurs on time scales extending from milliseconds to months, as well as on spatial scales ranging from subcellular and synaptic levels to those of networks and whole-brain structures. 

\newpage
\tableofcontents

\newpage
\section{Introduction}
Most animals share the ability to move in response to external stimuli, which results from having developed complex neural structures that allow for sophisticated processing of sensory information. From a phylogenetic perspective, as the nervous system changed from a simple net-like structure, such as in jellyfish (\textit{Cnidara}), to condensed ganglia and centralized brains -- starting with flat interstitial worms (\textit{Acoelomorpha}) up to mammals (\textit{Chordata})--, a new cell type could be recognized in morphological studies: glial cells \citep{Hartline_Glia2011,Laming_NBR2000}. The importance of this new cell type for a functional brain is arguably reflected by the increase in the relative number of glial cells during evolution: there are roughly equal numbers of glial cells and neurons in mammals \citep{Herculano_Glia2014} whereas, by contrast, glia constitutes only 10\% of invertebrate neural cells in \textit{Drosophila} or \textit{C. elegans} \citep{Hartline_Glia2011,Bullock_Freeman1965}.

It is currently unknown whether glial cells from different clades share similar functions: the molecular, morphological, and physiological identity of glial cells across and within clades indeed remains a topic of active investigation \citep{Hartline_Glia2011}. In the mammalian brain, the most numerous glial cells appear to fall into two categories: (i)~macroglia, which is prominently constituted of oligodendrocytes and astrocytes, and (ii)~microglia \citep{Lawson_Neurosci1990,Pelvig_NA2008}. Significantly, despite their different developmental origin -- macroglia originating from the embryonic ectoderm while microglia coming from the mesoderm --, both cell types share a common feature: that is physical proximity to neurons by an elaborated branched anatomy that interweaves with neural processes \citep{KettenmannRansom_Book2013}. This feature can be regarded as the teleological evidence of the purpose of glia for being an integral part of both the structure and function of neural networks of the mammalian brain.

Oligodendrocytes, for example, are responsible for myelination of axons, which is essential for their trophic support to attain long lengths while allowing for evolved organisms, like mammals and vertebrates in general, to achieve great sizes \citep{Nave_Nature2010}. At the same time, myelination provides axons with high membrane electrical resistance and low capacitance, which prevents current loss and enables rapid and efficient conduction of action potentials -- a prerogative of complex nervous systems to operate quickly and efficiently \citep{Zalc_Science2000}. As nervous systems increase in complexity, and this complexity correlates with the increasing sophistication of neurological function, a trend in increased complexity is also observed for astrocytes \citep{Verkhratsky_PTRSB2016}. Humans and primates indeed possess astrocytes that are larger and more branched than rodent astrocytes, with humans astrocytes generally being the largest \citep{Oberheim_JN2009}. In agreement with this trend, if we graft human astrocytes in mice, we get animals that show enhanced learning and memory capabilities \citep{Han_CSC2013}. Finally, microglia, too, can be framed within the evolutionary perspective, in connection with the appearance of compact neural masses and the increased demand thereby of specialized phagocytic and immune functions, including mechanisms of neural protection \citep{Posfai_Neurosci2019}. Remarkably, microglia invades the neural tube during embryogenesis before epithelial cells, neurons, and macroglia, being in the ideal position to regulate angiogenesis as well as neuro- and gliogenesis. It is now emerging that such regulation could extend beyond embryonic life as microglia can promote the formation and dismantling of functional neuron-glial networks by axon reorganization during development and by regulation of genesis and pruning of synapses in the mature brain \citep{Posfai_Neurosci2019}.

Although lagging behind classical computational neuroscience, theoretical and computational approaches are beginning to emerge to characterize different aspects of neuron-glial interactions \citep{DePitta_Springer2019}. This chapter aims to provide essential knowledge on neuron-glial interactions in the mammalian brain, leveraging on computational studies that focus on structure (anatomy) and function (physiology) of such interactions in the healthy brain. Although our understanding of the need of neuron-glial interactions in the brain is still at its infancy, being mostly based on predictions that await for experimental validation, simple general modeling arguments borrowed from control theory are introduced to support the importance of including such interactions in traditional neuron-based modeling paradigms.

\section{Anatomy of neuron-glial interactions}
\subsection{Myelination of axons by oligodendrocytes}
Myelination by oligodendrocytes and axon growth seem to be entirely interdependent since there are direct relationships among the age at which axons are myelinated, their final diameter in the adult, and the development of the associated oligodendrocytes \citep{Butt_2013Ch}. Accordingly, oligodendrocytes and their axons should be considered together as functional units rather than separate functional entities. Myelinated axons can thus be thought to be functionally specialized since internodes provided by myelinic ensheathing, and nodes of Ranvier develop in a concerted fashion to fulfill specific computational tasks \citep{Tomassy_TiCB2016,Hughes_NN2018}. 

Electrical cable theory \citep{FitzHugh_BJ1962,Goldman_BJ1968,Moore_BJ1978,Waxman_MN1980,Richardson_MBEC2000} predicts that regulation of myelin thickness, internode length and node geometry by oligodendrocytes  \citep{Ullen_NGB2009,Fields_TiNS2008,Arancibia_eLife2017} could account for fine-tuning of action potential~(AP) shape and conduction velocity
\citep{Kimura_FNA2009,Tomassy_Science2014,Ford_NC2015,Hughes_NN2018} with important functional consequences. For instance, variations of internodal length associated with axon diameter dictate the extent of co-activation of nodal \ce{Na^+} with low-threshold \ce{K^+} channels. In turn, a triaxial cable model of myelinated axons in sound processing circuits reveals how the synchronous activation of these channels may account for large, fast-rising and brief~APs, differentiating axons that can propagate~APs at high rates, and that belong to neurons that encode for high-frequency sounds, from axons that cannot, and instead depart from neurons that respond to low-frequency sounds \citep{Halter_JTB1991,Ford_NC2015}. In those circuits, adjustments of internodal length may also occur in close register with nodal diameter, crucially setting the timing of AP~arrival at synaptic terminals, which, in turn, underpins precise binaural processing of temporal information for sound localization \citep{Carr_JN1990,Carr_CON2001,Carr_BBE2002,McAlpine_TiNS2003,Seidl_JN2010,Ford_NC2015}. Together with delays induced by the~AP-generation dynamics \citep{Fourcaud_JN2003} and those rising from synaptic processing \citep{Markram_JP1997}, axonal conduction delays due to myelination by oligodendrocytes are an important property of neural interactions, which may induce a wealth of dynamical states with different spatiotemporal properties and domains of multistability that could serve different computational purposes \citep{Roxin_PRL2005,Roxin_PD2011}.

\subsection{\Muller\ glia in the retina}
Formation of axon-myelin units by oligodendrocytes to fulfill specific functional tasks is only one of the possibly many cases of how neuronal structures and glial cells' morphology develop in a tight association. Another example is the retina of the vertebrate eye. Notably, this structure is inverted with respect to its optical function so that light must pass through several tissue layers before reaching the light-detecting photoreceptor cells. In doing so, projected images on the retina would rapidly deteriorate due to light scattering by optical and geometrical inhomogeneities of those different layers if it were not for specific properties and arrangement of structures and cell assemblies of the retina that ensure correct vision \citep{Tuchin_B2000,Masland_NN2001}. In this scenario, the most common type of retinal glial cells, known as \Muller\ cells, show a characteristic cylindrical, funnel-like shape that spans the entire thickness of the retina and functions as the waveguide of visible light from the retinal surface to the photoreceptor cell layer \citep{Franze_PNAS2007}. Numerical solution of the Helmholtz equation in a model of retinal tissue suggests that this property of \Muller\ cells to guide light crucially enhances vision acuity since, like fiber optics that penetrate through the retinal layers, these cells can propagate light straight down to the photoreceptor cell they associate with \citep{Labin_PRL2010}. In particular, every mammalian \Muller\ cell is coupled, on average, to one cone photoreceptor cell, that is responsible for sharp seeing under daylight conditions (i.e., photopic vision), plus several rod photoreceptor cells, serving low-light-level, night (scotopic) vision \citep{Reichenbach_PRER1995}. Remarkably, the spectrum of light transmitted through \Muller\ cells, derived from theoretical calculations based on data of the human parafoveal retina, matches almost perfectly the absorption spectra of the medium- and long-wavelength human cone photoreceptors, predicting a gain in photon absorption by a factor of~$\sim$7.5 by these photoreceptors, and by a factor of~$\sim$4 by human short-wavelength cones \citep{Labin_NC2014}. At the same time, the spectrum of light leaking outside \Muller\ cells matches the absorption spectrum of human rod photoreceptors. In this fashion, \Muller\ cells provide a mechanism to improve cone-mediated photopic vision, with minimal interference with rod-mediated scotopic vision since not only they guide light of relevant wavelengths for cone visual pigments directly towards cones, but also they allow for light of wavelengths more suitable for rod vision to leak to surrounding rods \citep{Labin_NC2014}.

\subsection{Astrocytes and regulation of volume transmission}
Broadly speaking, the factors underpinning morphogenesis of neuronal structures in tight association with glial cells remain poorly understood, although they are thought to encompass a combination of transcriptional programs and a battery of molecular signals that seem to be regulated both developmentally and regionally, in a cell-specific manner \citep{Jadhav_PRER2009,Eroglu_Nature2010,Gotz_2013Ch,Stassart_2013Ch}. A paramount example of this complex combination of factors is provided by astrocytes, which are glial cells predominantly found in the cortex and hippocampus and are recognized as critical regulators of the neuronal connectome in those brain regions \citep{Fields_Neuron2015}. In the postnatal, developing brain, astrocytes are indeed found in spatially distinct domains and express domain-specific genes that are needed to support the formation of specific neural circuits and neuronal subtypes \citep{Molofsky_Nature2014}. On the other hand, they also secrete a variety of molecules that regulate with spatiotemporal specificity all stages of the genesis of functional neural circuits, such as the establishment of immature (silent) synapses between axon and dendrites, the conversion of these silent synapses into active ones by insertion (and maintenance) of functional receptors, and the elimination of excess synapses and synaptic pruning to refine connections in neural circuits \citep{Allen_CON2013,Clarke_NN2013}. Remarkably, this close relationship between astrocytes and synapses continues in the adult brain, with the processes of astrocytes that infiltrate into the neuropil and wrap themselves around synapses in a seemingly non-random fashion \citep{Reichenbach_BRR2010}, engaging in a multitude of signals (\figref{fig:interactions}), with figures approaching~20--180 thousand synapses contacted by a single astrocyte in the rodent brain \citep{Bushong_JN2002}, but $\sim$270~thousand to 2~million synapses per astrocyte in the human brain \citep{Oberheim_JN2009}. What are the possible functional purposes of this ensheathing?

One obvious possibility is that processes of astrocytes are strategically positioned at synaptic loci to act as physical barriers that constrain and regulate extracellular diffusion of neurotransmitters \citep{Ventura_JN1999}. With this regard, a realistic reconstruction of the extracellular space~(ECS) in between astrocytic processes and glutamatergic synapses of the hippocampus -- one of the brain structures where the morphological association of astrocytes with synapses has been best-characterized \citep{Seifert_CTR2017} -- revealed a strong anisotropy for glutamate diffusivity, directly imputable to the peculiar arrangement of astrocytes with the surrounding neuropil \citep{Kinney_JCN2013}. This anisotropy ensues from the nonuniform width of~ECS, which forms thin sheets in the periastrocytic space and grows wider into tunnels in correspondence with synapses and axons. Monte Carlo diffusion simulations show that the diffusion rate of glutamate through the neuropil critically depends on the proportion of sheets~vs. tunnels, since the smaller median extracellular width of astrocyte-delimited sheets, poses a diffusion barrier to glutamate molecules, corralling them into tunnels which favor their diffusion instead \citep{Kinney_JCN2013}. This scenario, along with the expression of glutamate transporters by perisynaptic astrocytic processes \citep{Danbolt_Book2002}, would ultimately provide a mechanism to canalize glutamate to specific targets while keeping low the risk of excitotoxicity brought forth by the fact that glutamate is not normally destroyed in the~ECS \citep{Rothstein_Neuron1996}.

The degree of astrocytic ensheathing at glutamatergic synapses indeed directly dictates the rate of glutamate uptake by glial transporters -- the primary mechanism of glutamate uptake in the adult brain \citep{Danbolt_PN2001} --, regulating the time course of this neurotransmitter in the synaptic cleft, and its potential to spill out to neighboring synapses \citep{Clements_TiNS1996}. Remarkably, the degree of astrocyte ensheathment seems inversely correlated with the size of dendritic spines \citep{Medvedev_PTRSB2014}, suggesting that, in agreement with independent theoretical arguments \citep{Barbour_JN2001}, smaller synapses are best sealed by astrocytic processes to minimize glutamate spillover. On the contrary, glutamate spillover could occur at larger synapses, for which the degree of astrocytic ensheathing is only up to approximately one-third of their perimeter \citep{Ventura_JN1999}. To the extent that thin-spine~vs. large-spine synapses can respectively be assimilated to silent~vs. functional synapses \citep{Matsuzaki_NN2001}, the latter possibility hints a crucial role of astrocytic ensheathing in the regulation of glutamate clearance at functional excitatory synapses, with the potential to regulate independent synaptic activation~vs. synaptic cross-talk \citep{Huang_CON2004}. 

\section{Physiology of neuron-glial interactions}
\subsection{General modeling framework}
If the reciprocal disposition of neurons with respect to glial cells in the brain fulfills essential anatomical constraints for their interaction, the existence of two-way dynamic signaling, from neurons to glia and vice verse, accounts for the physiology of this interaction. With this regard, a common framework to model networks of interacting neurons \citep{ErmentroutTerman_2010} may be borrowed to illustrate essential functional consequences of neuron-glial interactions. Denoting by \xn\ and \gs\ the activities of neuronal and glial interacting ensembles, respectively defined as column vectors in $\mathbb{R}^N$ and $\mathbb{R}^G$, then the glial activity can generally be described by 
\begin{align}
\tau_\upgamma \dot{\gs} &= -\gs + \mathbf{\mathcal{F}}_\upgamma\rpar{\mathbf{b}_\upgamma + \mathbf{I}_\upgamma\rpar{\xn,\gs}} \label{eq:glia-activity}
\end{align}
where $\mathbf{\mathcal{F}}_\upgamma : \mathbb{R}^g \rightarrow \mathbb{R}^G$ ($g\gtreqless G$) represents the glial input-output transfer function. The argument of this function -- that is the input to glia -- consists of (i)~an external bias $\mathbf{b}_\upgamma \in \mathbb{R}^g$ coming, for example, from brain areas away from the neuron-glial ensemble under consideration, and of (ii)~a recurrent input $\mathbf{I}_\upgamma\rpar{\xn,\gs} : \mathbb{R}^{N+G}\rightarrow \mathbb{R}^g$ due to neuronal and glial activities within the ensemble itself. The physiological correlate of $\gs$, along with the choice of $\mathbf{\mathcal{F}}_\upgamma$ and $\mathbf{I}_\upgamma$ depend on what ``glial activity'' refers to, and it will be clarified as follows. Similarly, without specifying for now what ``neuronal activity'' is, an equation analogous to~\ref{eq:glia-activity} may be written for \xn, i.e.
\begin{align}
\tau_\mathrm{x} \dot{\xn} &= -\xn + \mathbf{\mathcal{F}}_\mathrm{x}\rpar{\mathbf{b}_\mathrm{x} + \mathbf{I}_\mathrm{x}\rpar{\xn,\gs}} \label{eq:neuron-activity}
\end{align}
\noindent
where $\mathbf{\mathcal{F}}_\mathrm{x} : \mathbb{R}^n \rightarrow \mathbb{R}^N$, $\mathbf{b}_\mathrm{x} \in \mathbb{R}^n$ and $\mathbf{I}_\mathrm{x}\rpar{\xn,\gs} : \mathbb{R}^{N+G}\rightarrow \mathbb{R}^n$ ($n\gtreqless N$) . In the previous equations, what the ``activities'' \xn\ and \gs\ refer to are suitable measurables of neuronal and glial physiology, respectively. Membrane voltage, firing rate, and synaptically-released neurotransmitters are typical examples of such measurables for neurons. For glia instead, cytosolic calcium is probably the most hitherto studied readout of these cells in response to neuronal stimulation, yet possibly not the only one, since other intracellular ions like \Na, \K, \H, \Cl\ or \HB, along with a whole cassette of molecules involved in the genesis of calcium signaling, could also be indicators of glial stimulation \citep{DePitta_Springer2019_Ch1}. At the same time, the notion of neuronal and glial ensembles is somehow broad. We may encompass by this notion either brain structures, including whole networks of neurons and glia or components of these networks, for example, myelinated axons with their oligodendrocytes, or groups of synapses with their perisynaptic glial processes (of astrocytic or microglial origin). In this context, the emerging notion that calcium homeostasis, as well as the homeostasis of other ions and macromolecules, could be compartmented within the glial cell's morphology \citep[][see also \see{Computational Modeling of Intracellular Astrocyte \ca\ Signals}]{Breslin_PCB2018}, allows using equations akin to~\ref{eq:glia-activity} and~\ref{eq:neuron-activity} to model dynamics of neuron-glial interactions at multiple spatial scales, where scale-specific aspects of such interactions may be lumped into the choice of the transfer functions $\mathbf{\mathcal{F}}_i$ (with $i=\xn,\,\gs$), their input arguments $\mathbf{b}_i,\, \mathbf{I}_i$, and ultimately, the components of the activity vectors \xn\ and \gs.

The functions $\mathbf{\mathcal{F}}_i$ and $\mathbf{I}_i$ are generally nonlinear with respect to their input variables. This hinders any attempt to look beyond the general form of equations~\ref{eq:glia-activity} and~\ref{eq:neuron-activity} to fathom neuron-glial interactions with the aim to derive different mechanisms of action and function that could be shared (or not) by different pathways of interaction. The situation greatly simplifies if we limit our analysis to the first order expansion of $\mathbf{I}_i$ around the resting point $(\xn_0,\gs_0)$. In this case, equations~\ref{eq:glia-activity} and~\ref{eq:neuron-activity} become
\begin{align}
\tau_\upgamma \dot{\gs} &\approx -\gs + \mathbf{\mathcal{F}}_\upgamma\rpar{\mathbf{b}_{0_\upgamma} + \mathbf{W}_\mathrm{x} \rpar{\xn -\xn_0} + \mathbf{J}_\upgamma \rpar{\gs - \gs_0}} \label{eq:glia-activity-1st}\\
\tau_\mathrm{x} \dot{\xn} &\approx -\xn + \mathbf{\mathcal{F}}_\mathrm{x}\rpar{\mathbf{b}_{0_\mathrm{x}} + \mathbf{J}_\mathrm{x} \rpar{\xn -\xn_0} + \mathbf{W}_\upgamma \rpar{\gs - \gs_0}} \label{eq:neuron-activity-1st}
\end{align}
where, for $i=\xn,\,\gs$, $\mathbf{b}_{0_i} = \mathbf{b}_i + \mathbf{I}_i(\xn_0,\gs_0)$, $\mathbf{J}_i = \rpar{D_i \mathbf{I}_i(i,j)}_{(\xn_0,\gs_0)}$ and $\mathbf{W}_i = \rpar{D_j \mathbf{I}_i(i,j)}_{(\xn_0,\gs_0)}$, with $D_\mathrm{x}\mathbf{I}_i = \rpar{\frac{\partial \mathbf{I}_i}{\partial x_0},\ldots,\frac{\partial \mathbf{I}_i}{\partial x_{N-1}}}$ and $D_\upgamma\mathbf{I}_i = \rpar{\frac{\partial \mathbf{I}_i}{\partial \upgamma_0},\ldots,\frac{\partial \mathbf{I}_i}{\partial \upgamma_{G-1}}}$. In this way, it is possible to distinguish between different contributions to the evolution of neuronal (glial) activity by $\mathbf{\mathcal{F}}_\mathrm{x}$ ($\mathbf{\mathcal{F}}_\upgamma$), in particular separating those that are due to neuronal (glial) elements weighted by the matrix $\mathbf{J}_\mathrm{x}$ ($\mathbf{J}_\upgamma$) from those that ensue instead from glial (neuronal) elements and are weighted by the matrix $\mathbf{W}_\mathrm{x}$ ($\mathbf{W}_\upgamma$). Because matrices $\mathbf{J}_\mathrm{x},\,\mathbf{J}_\upgamma$ lump couplings of neuronal (respectively glial) activities with themselves, whereas matrices $\mathbf{W}_\mathrm{x},\,\mathbf{W}_\upgamma$ couple neuronal activities with glial activities and vice verse, we hereafter refer to them as \textit{homotypic} and \textit{heterotypic} weight (or connection) matrices, respectively. 

\begin{figure}[!tp]
\centering
\includegraphics[width=\textwidth]{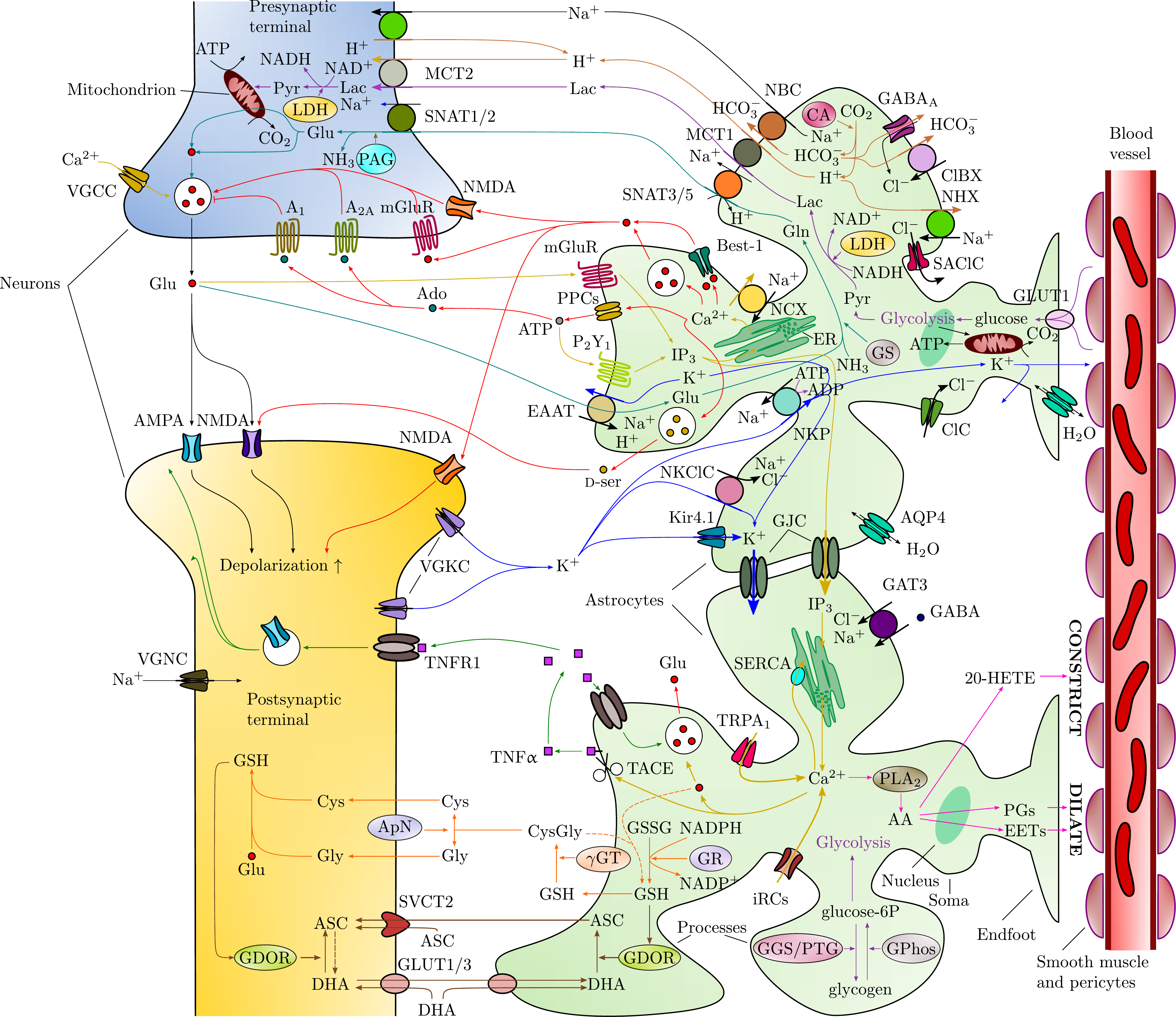}
\caption{Common interactions between astrocytes and glutamatergic synapses.}
\label{fig:interactions}
\end{figure}

\begin{figure}[!tp]
\centering
\includegraphics[width=\textwidth]{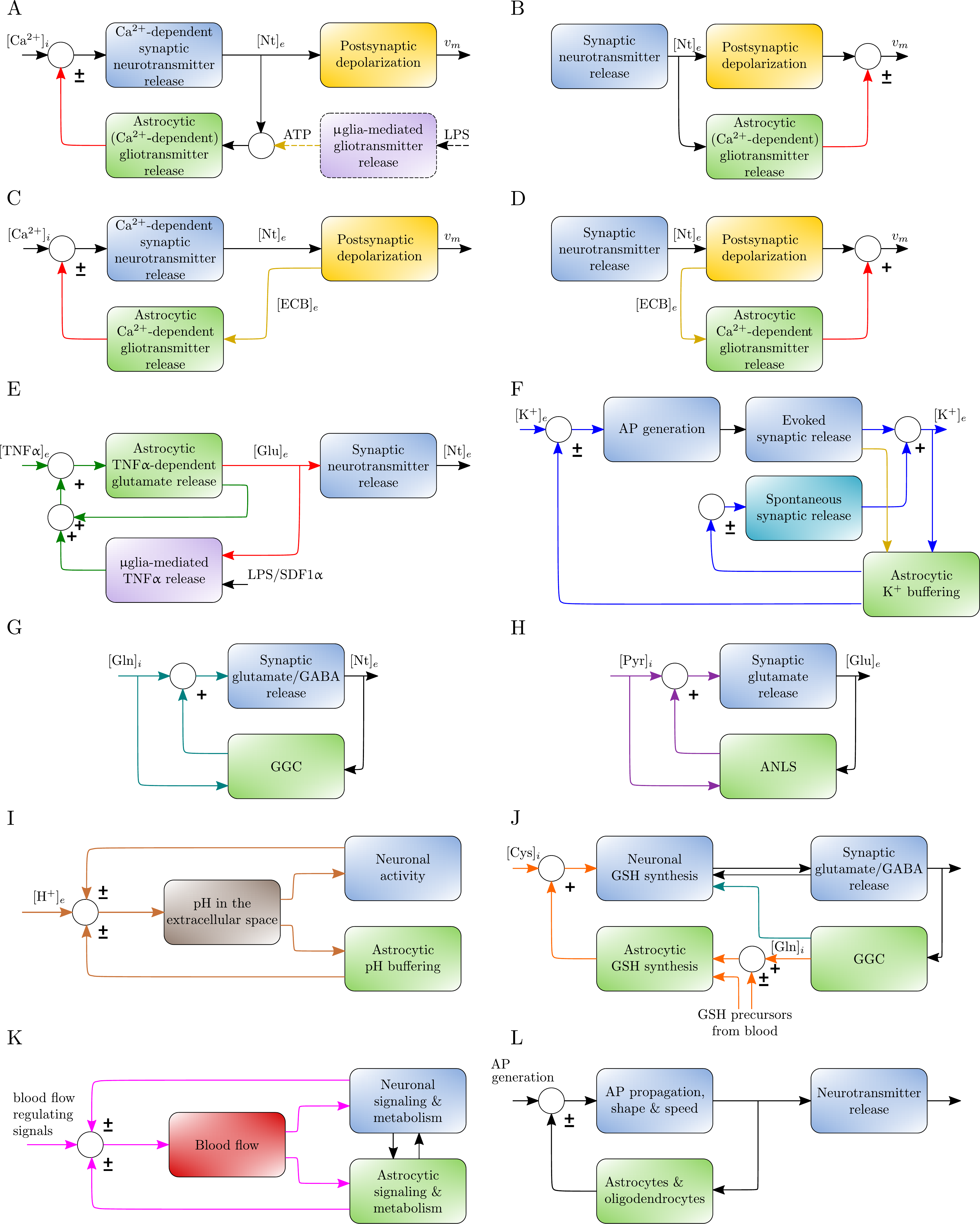}
\caption{Feedback and feedforward pathways in neuron-glial interactions.}
\label{fig:schemes}
\end{figure}

The input terms associated with heterotypic connections represent the actual interaction pathways, for signaling from neurons to glia ($\mathbf{W}_\mathrm{x} \xn$), and vice verse, from glia to neurons ($\mathbf{W}_\upgamma \gs$). In several practical scenarios (\figref{fig:schemes}), this signaling is mediated by specific molecules or ions, that may themselves be regarded as measures of neuronal (glial) activity, or can alternatively be expressed in terms of \xn\ (respectively \gs). Consider for example synaptically-evoked gliotransmission \citep[][see also \textit{yellow} and \textit{red} pathways in \figref{fig:interactions}]{Araque_Neuron2014}. This form of neuron-glial interaction consists of the activity-dependent release of neuroactive molecules from astrocytes, including neurotransmitters like glutamate, ATP, GABA, or \dserine\ that, for their glial origin, are termed ``gliotransmitters.'' The release can be triggered by synaptically-released neurotransmitters and be mediated by different intracellular pathways that often associate with transient calcium signaling in the astrocyte \citep{Savtchouk_JN2018}. In turn, gliotransmitters can modulate synaptic transmission both presynaptically -- binding to receptors that regulate the probability of release of neurotransmitter-containing synaptic vesicles --, and postsynaptically, gating receptors/ion channels that control neuronal depolarization (and excitability). In this description, synaptically-released neurotransmitters mediate neuron-to-astrocyte signaling, whereas gliotransmitters are responsible for the astrocyte-to-neuron signaling. To couple these two signaling pathways by the above framework, however, biophysical correlates for their functional interdependence must also be explored. For the dependence of gliotransmitter release on synaptic release, at least at first approximation, gliotransmitter release is expected to rise with synaptic neurotransmitter release \citep{Araque_Neuron2014}. For the dependence of this latter on gliotransmitter release instead, the reasoning is as follows: (i)~activation of presynaptic receptors by gliotransmitters increases with gliotransmitter release; in turn (ii)~modulation of intrasynaptic \ca, which is linked with $p$ -- the synaptic release probability (\see{Calcium-Dependent Exocytosis, Biophysical Models}) --, correlates with the degree of receptor activation by gliotransmitters; (iii)~synaptic release is proportional to $p$. In this fashion, the simplest scenario sees $\gs$ in the above equations reducing to a scalar variable that is associated with the fraction of gliotransmitter-activated receptors, whereas $p$ replaces $\xn$. Then, assuming that at rest $\gamma_0=0,\,p_0=p^*>0$, the simplest model for the presynaptic pathway of synaptically-evoked gliotransmission (\figref{fig:schemes}A) derived by equations~\ref{eq:glia-activity-1st} and~\ref{eq:neuron-activity-1st} reads 
\begin{align}
\tau_\gamma \dot{\gamma} &= -\gamma + \mathcal{F}_\gamma \rpar{b_{\gamma_0} + W_\gamma p + J_\gamma \gamma}\label{eq:gt-gamma}\\
\tau_p \dot{p}     &= p^*-p + \mathcal{F}_p \rpar{(1-J_p)p^* + J_p p + W_p\gamma}\label{eq:gt-p}
\end{align}
where $b_{\gamma_0}$ may or not be present, depending on the presence or not of other mechanisms triggering gliotransmitter release that are independent of stimulation of the very synapse modulated by gliotransmission, e.g., microglia-mediated \ca\ signaling in the astrocyte \citep{Pascual_PNAS2011}. The homotypic weight $J_\gamma$ accounts for the theoretically-predicted phenomenon of gliotransmitter-triggered gliotransmission \citep{LarterCraig_Chaos2005}, whereas the homotypic weight $J_p$ can generally be accounted by mechanisms of short-term plasticity (\see{Short Term Plasticity, Biophysical Models}). With regard to the heterotypic weights instead, it generally is $W_\gamma > 0$ regardless of the nature of synaptic neurotransmitters, that is both excitatory and inhibitory neurotransmitters are excitatory for glial activation \citep{Durkee_Glia2019}. The sign of $W_p$ instead is dependent on the nature of gliotransmission, which can either decrease or increase synaptic release \citep{DePitta_Neurosci2015}. Remarkably, while the sign of $W_p$ is set by functional and anatomical features of synapse-astrocyte ensembles \citep{DePitta_Springer2019_Ch10}, the further possibility that the polarity of gliotransmission could also depend on specific patterns of activity \citep{Covelo_eLife2018} can be accounted by choice of $\mathcal{F}_p$.

If we aim for a more general understanding of gliotransmission on synaptic transmission, it is then useful to think of the efficacy (or strength) of synaptic transmission in terms of the product $w=p\cdot q$, where $q$ is the probability of activation of postsynaptic receptors by the presynaptically released neurotransmitter. In this context then, choosing $\xn = \rpar{p,q}^T$ with $q_0 = q^*$, results in the system of three equations in the general form of
\begin{align}
\tau_\gamma \dot{\gamma} &= -\gamma + \mathcal{F}_\gamma \big(b_{\gamma_0} + W_\gamma^{(p)} p + W_\gamma^{(q)} q + J_\gamma \gamma \big)\label{eq:gt2-gamma}\\
\tau_p \dot{p}     &= p^*-p + \mathcal{F}_p \big((1-J_p)p^* - W_p^{(q)}q^* + J_p p + W_p^{(q)}q + W_p^{(\gamma)}\gamma \big)\label{eq:gt2-p}\\
\tau_q \dot{q}     &= q^*-q + \mathcal{F}_q \big(- W_q^{(p)}p^* + (1-J_q)q^* + W_q^{(p)}p + J_q q + W_q^{(\gamma)}\gamma \big)\label{eq:gt2-q}
\end{align}
\noindent
The coupling factors $J_q,\,W_\gamma^{(q)},\,W_p^{(q)}$ and $W_p^{(\gamma)}$ ensue from the consideration of endocannabinoid signaling, which may retrogradely suppress synaptic activity while, at the same time, trigger \ca-dependent gliotransmitter release from the astrocyte \citep{Navarrete_PTRSB2014}. In particular, because this signaling pathways depends on postsynaptic depolarization ($v_m$), and this latter is dependent on incoming synaptic activity by $w$, it is safe to assume that endocannabinoid signaling can be expressed in terms of $p$ and $q$ by a proper choice of the transfer functions $\mathcal{F}_\gamma,\, \mathcal{F}_p,\,\mathcal{F}_q$ and their input arguments $I_\gamma,\,I_p,\,I_q$ in the original equations~\ref{eq:glia-activity} and~\ref{eq:neuron-activity} (Figures~\ref{fig:schemes}C,D). On the other hand, when no retrogade signaling (including endocannabinoid) is taken into account, then $J_q=0$ and also $W_\gamma^{(q)}=W_p^{(q)}=W_p^{(\gamma)}=0$, allowing for a significant simplification of the above equations. Even more so if the analysis is restricted to the sole postsynaptic pathway of gliotransmission without considering any mechanism of short-term plasticity, for which, synaptic dynamics reduces to $\tau_p \dot{p} = p^*-p$ and $\tau_q \dot{q} = q^*-q + \mathcal{F}_q \big(- W_q^{(p)}p^* + q^* + W_q^{(p)}p + W_q^{(\gamma)}\gamma \big)$ with $\gamma$ being described by an equation like~\ref{eq:gt-gamma}. In this scenario gliotransmission release ($\gamma$) becomes uncoupled from $q$ (but not the opposite), and the dynamics of both $\gamma$ and $q$ is driven by $p$ (\figref{fig:schemes}B). The resulting $q(t)$ can essentially be expressed in terms of $p(t)$ but, differently from a standalone synapse, the effect of $p$ on $q$ (and thus on $v_m$) also incorporates, possibly in a nonlinear fashion, the effect of gliotransmission. Remarkably, this scenario could underpin several examples of neuron-glial interactions such as: (i)~regulation of long-term synaptic plasticity by \dserine\ released from astrocytes in the hippocampus, hypothalamus and the cortex \citep{BainsOliet_TINS2007}; (ii)~AMPA receptor internalization in the cerebellum by \dserine\ originating from Bergmann glial cells \citep{Kakegawa_NN2011}; (iii)~AMPA receptor insertion by adrenergic stimulation of~ATP release from hypothalamic astrocytes \citep{Gordon_NN2005}; (iv)~modulation of~AMPA and~GABA receptor trafficking by astrocytic \tnfa\ \citep{Stellwagen_JN2005}; and (v)~slow inward and outward currents (SICs and SOCs) respectively mediated by glutamate and~GABA from astrocytes \citep{Fellin_Neuron2004,LeMeur_FCN2012}. 

The distinction between homotypic vs. heterotypic couplings brought forth by equations~\ref{eq:glia-activity-1st} and~\ref{eq:neuron-activity-1st} allows the analysis of otherwise complex neuron-glial interactions in terms of combinations (to the first-order approximation) of feedback and feedforward mechanisms, which can conveniently be illustrated by functional diagrams such as those in \figref{fig:schemes}. Furthermore, consideration of these diagrams can help classify different pathways of interaction between neurons and glia based on their coupling mechanisms. In this fashion, activity-dependent presynaptic pathways of gliotransmission can all be assimilated to a  feedback mechanism on synaptic release, which can be either positive or negative, depending, as previously stated, on the nature of astrocyte-synapse coupling and synaptic dynamics (\figref{fig:schemes}A). Conversely, the postsynaptic pathway of gliotransmission is tantamount to a feedforward mechanism on postsynaptic membrane depolarization, which can also be positive or negative, depending on the gliotransmitter type and whether gliotransmission promotes receptor internalization or insertion (\figref{fig:schemes}B). Consideration of endocannabinoid signaling does not change the essence of these mechanisms, except for the fact that a further intermediate processing stage must be taken into account for the regulation of endocannabinoid release from postsynaptic terminals (\figref{fig:schemes}C). Nonetheless, differently from other pathways of synaptic transmission, in this latter case, postsynaptic pathways of gliotransmission may also feedback on endocannabinoid release since this process depends on postsynaptic depolarization (\figref{fig:schemes}D). On the other hand, glutamatergic gliotransmission could also contemplate the existence of positive feedback on the astrocyte by its release of the cytokine \tnfa\ (\figref{fig:schemes}E). Constitutive extracellular levels of this cytokine ([\tnfa]$_e$) may indeed promote glutamate release from astrocytes but can also regulate the production of \tnfa\ by these cells \citep{Santello_Neuron2011}. Furthermore, microglia could also regulate glutamatergic gliotransmission, releasing ATP as part of their immune response, and thereby promoting \ca-dependent glutamate release from astrocytes \citep{Pascual_PNAS2011}. Alternatively, extracellular glutamate ($[\ce{Glu}]_e$) may trigger the release of \tnfa\ from microglia, which could stimulate the further release of glutamate from astrocytes in, yet another, positive feedback fashion \citep{Bezzi_NN2001}.

Potassium ion homeostasis is also involved in multiple feedback pathways mediated by glia (\figref{fig:schemes}F). Neuronal activity results in local accumulations of extracellular \K\ ($[\K]_e$) that may promptly be abated by spatial buffering by glia \citep{Kofuji_Neurosci2004}. Significantly, synaptically-evoked \ca\ signaling in astrocytes (\textit{yellow pathway}) can also regulate $[\K]_e$ to modulate both spontaneous synaptic release at excitatory synapses and neuronal firing \citep{Wang_NN2006}. In the cerebellum, \ca-dependent \K\ uptake by Bergman glia can dramatically alter the firing dynamics of Purkinje cells \citep{Wang_PNAS2012}. At cortical axons instead (\figref{fig:schemes}L), depolarization of myelinating oligodendrocytes following AP~conduction is speculated \citep{Debanne_PR2011} to alter extracellular levels of \K\ and \Na\ locally, at nodes of Ranvier, thereby transiently promoting action potential generation which could speed up AP~conduction by~10\% \citep{Yamazaki_NGB2007}. On the other hand, astrocytes, too, could influence AP~generation at nodes of Ranvier, specifically by increasing AP~duration by \ca\-dependent glutamate release, although the activity requirements for this pathway are not known \citep{Sasaki_Science2011}.

Very often, a signaling pathway or a biophysical process may be shared by different mechanisms of neuron-glial interactions. Extracellular glutamate and its uptake by astrocytes, for example, are key components of the so-called glutamate--glutamine cycle (GCC), as well as of the astrocyte-to-neuron lactate shuttle (ANLS) (see \see{Brain Energy Metabolism}). In the~GCC (\figref{fig:schemes}G), glutamate taken up by astrocytic transporters is converted into glutamine, which is transported back to neurons, where it is presumably reconverted to glutamate, thereby guaranteeing the supply of this neurotransmitter to sustain synaptic transmission \citep{Xiang_NCI2003}. In the ANLS instead (\figref{fig:schemes}H), astrocytic lactate -- ensuing from ATP demand to counteract intracellular \Na\ accumulation by glutamate uptake --, is shuttled to neurons where it is converted to pyruvate, which is necessary for ATP synthesis by mitochondria, and thus it mediates a positive feedback loop that can sustain the energetic cost of prolonged neuronal firing activity \citep{Allaman_TiNS2011}. Remarkably, both ANLS and GCC could also be integral components of other pathways of neuron-glial interactions. Lactate, for example, may also cover synaptic energy demand indirectly by triggering prostaglandin release from astrocytes, which, in turn, increases blood flow and thereby, glucose supply to neurons \citep{Attwell_Nature2010}. Alternatively, it could also operate as gliotransmitter \citep{Tang_NC2014}. Glutamine produced by~GCC, on the other hand, could also be a precursor of glutathione (GSH) in the synthesis of this latter by astrocytes (\figref{fig:schemes}J), which is a key process in neuroprotection by glia. Neurons are also capable of synthesizing~GSH from glutamate, but this is constrained by their intracellular availability of cysteine and glycine. Astrocytes can, however, backup on the availability of these two amino acids in neurons by releasing GSH, which is extracellularly cleaved to produce cysteinylglycine that neurons, in turn, cleave to supply cysteine and glycine to their intracellular resources \citep{Dringen_PN2000}. Finally glutathione metabolism could also be linked with vitamin~C (ascorbic acid) exchange between neurons and glia as illustrated by the \textit{brown pathway} in \figref{fig:interactions} \citep{Castro_JNC2009,May_2012}.

In several cases, neuron-glial interactions add to existing neuronal feedback mechanisms. Two examples, respectively, are the control of extracellular~pH and the regulation of blood flow by astrocytes. Neuronal electrical activity results in extracellular alkalinization and intracellular acidification around neuronal cell walls, which can considerably alter neuronal excitability and synaptic currents \citep{Deitmer_PN1996}. This is put in place by a multitude of diverse mechanisms that rely on physiochemical buffers, metabolic reactions, and membrane transport systems within neurons and in the interstitial fluid that make extracellular and intracellular neuronal pH completely interdependent. On the other hand, intracellular pH of astrocytes is also dependent on extracellular pH, and astrocytes, akin to neurons, express a variety of transport and buffering mechanisms that make them capable of influencing \H\ homeostasis locally and brain-wide actively. By similar arguments, both neurons and glial cells require a constant supply of metabolites and chemical species by the bloodstream. Yet, they can regulate blood flow by the independent release of vasoactive signals \citep{Attwell_Nature2010}, as well as do so in interaction with each other by the other neuron-astrocyte interaction pathways hitherto presented.

The existence of feedback and feedforward pathways of interactions between neuronal and glial elements challenges traditional connectomics. We come to realize in fact that neurons are not merely connected by synapses but also by extracellular glial-mediated pathways. Moreover, for some of these pathways, like at synapses where signal transmission is traditionally assumed to occur in one direction only -- from the presynaptic terminal to the postsynaptic one --, this directionality is replaced by a neuron-glial signaling network that is inherently recurrent. Recurrent connections are essential in network theory since they can control the stability of network activity and its robustness to noise. Accordingly, the model of an autaptic neuronal oscillator may conveniently be adopted to get a flavor of the potential relevance of glia-mediated synaptic recurrence on neuronal firing \citep{Volman_NC2007}. The autaptic oscillator consists of an excitatory neuron that exhibits periodic firing employing self-synapses (i.e., autapses) whose strength critically controls the neuron's firing rate and its stability \citep{Seung_JCN2000}. In this fashion, when all autapses also stimulate the same astrocytic domain and are equally affected by release-decreasing gliotransmission ensuing from it, activity-dependent fluctuations of neuronal firing rate emerge in phase opposition with astrocytic \ca\ spikes \citep{Volman_NC2007}. The mechanism would not be surprising \textit{per se} if it were not for the fact that it can also be observed in the presence of irregular bursting by the autaptic neuron. The theory of autaptic oscillators (without astrocyte) predicts in fact that the positive feedback exerted by excitatory autapses on the neuron only allows this latter to maintain its firing rate at the onset of stimulation, or to increase it, ultimately reaching unrealistic (epileptiform) firing rates \citep{Seung_JCN2000}. This scenario is avoided in the presence of release-decreasing gliotransmission, however, because any self-amplifying increase of neuronal activity that could be triggered, for example, by a burst of APs, promotes gliotransmitter release from the astrocyte, which promptly counteracts the neuron's runaway by excitation by depressing autaptic transmission. In this fashion, the negative feedback exerted by gliotransmission on synapses competes with the positive autaptic feedback to prevent the neuron from persistently generating APs at high rates. Rather, the opposite happens, as prolonged periods of depressed synaptic excitation, consistent with the long-lasting effect of release-decreasing gliotransmission observed \textit{in vitro} \citep{Araque_EJN1998}, substantially modify the firing statistics of the autaptic neuron, accounting for large interspike intervals. The ensuing neural bursting may accordingly be thought of as the alternation of APs with irregular periods of no activity and explains why the Gaussian distribution of interspike intervals otherwise observed for an autaptic neuron without astrocyte, is instead heavy-tailed in the presence of gliotransmission \citep{Volman_NC2007}.

\subsection{Modulation of neural excitability}
Two pathways for regulation of neuronal excitability by glia have hitherto been explored by computational studies: one by glia-mediated control of \K\ homeostasis, the other by modulation of excitatory synapses by astrocytes. In the first case, the fact that cortical astrocytes, M\"uller glia in the retina, or Bergmann glia in the cerebellum, buffer extracellular \K\ redistributing it to sites of lower concentration, locally changes the resting membrane potential of neurons by the \see{Goldman-Hodgkin-Katz Equation} as well as the Nernst potential of individual ions (\see{Nernst Equation}), thereby modulating the neuron's threshold for AP~generation. This modulation can be crucial to set the tone of neuronal firing in healthy physiological conditions since blocking glial buffering could turn random firing maintained by random external stimulation into periodic bursting, and eventually lead to permanent spike inactivation \citep{Bazhenov_JNP2004}. Mechanistically, glial \K\ buffering can be by different combinations of passive and active transport systems, which, however, share an interplay between glial ATP-driven \Na/\K\ pumps~(NKPs) and Kir~channels. Depending on neuronal activity, this interplay can mediate non-trivial positive feedback on neuronal firing, for example, promoting AP generation in conditions of low neuronal firing \citep{Somjen_JCN2008}, or turning regular spiking dynamics into bursting \citep{Somjen_JCN2008,Oyehaug_JCN2012,Cui_Nature2018}. 

Control of neuronal firing by modulation of excitatory synaptic transmission by astrocytes could instead occur, in principle, both by pre- and postsynaptic pathways of gliotransmission, and intrasynaptically, by glutamate uptake by astrocytic transporters. In the first scenario, the rationale for modulation essentially follows that of equations~\ref{eq:gt2-gamma}--\ref{eq:gt2-q}, where \ca-dependent gliotransmission could be triggered either by synaptically-activated astrocytic metabotropic receptors \citep{DePitta_NP2016} or by \Na\ influx into the astrocyte by glutamate transporters which in turn regulates \ca\ influx into the astrocyte by \Na/\ca-exchangers \citep{Flanagan_PCB2018}, though this latter possibility requires experimental validation. In the second scenario, detailed kinetic models of astrocytic transporters and~AMPA and~NMDA synaptic receptors suggest instead that control of glutamate clearance by astrocytic transporters could modulate postsynaptic receptors' activation and desensitization thereby modulating the kinetics of postsynaptic current. This results in subtle changes in spike arrival timings and spike failures that modulate the firing pattern of the postsynaptic neuron \citep{Allam_FCN2012}.

\subsection{Neuron-glial interactions in neural networks}
The importance of uptake of neurotransmitters such as glutamate and GABA by astrocytes in setting the tone of neural network activity has only been partially explored in the framework of \see{Neural Mass Action} modeling, yet producing two interesting predictions \citep{Garnier_JMN2016}. First is the observation that a deficiency of GABA uptake by astrocytes increases the threshold for neuronal activation in a linear fashion. Second is the prediction that, in the presence of deficient astrocytic glutamate uptake, neural activity may either be reduced or enhanced or may display a transient of high activity before stabilizing around a new regime whose firing frequency is close to the one measured in the absence of astrocyte deficiency. Significantly, the somehow counterintuitive possibility that neural activity could decrease despite extracellular glutamate accumulation due to deficits in astrocytic uptake may be observed only in the presence of sufficient interneuron hyperexcitability. 

More generally, in the framework of neuronal network theory, modulation of neuronal excitability and synaptic transmission by glia is expected to modulate the balance of excitation~(E) vs. inhibition~(I) with the potential to regulate network activity dramatically. This possibility follows by the analysis of the few available neuron-glial network models, which come in different flavors in terms of the combination of different E--I network configurations and different choices of neuronal (and synaptic) models and astrocytic signaling pathways \citep{Savin_JRSI2009,Ullah_JCN2009,Volman_PCB2013,Savtchenko_PTRSB2014}. Nonetheless, all these models eventually envisage an effect of glial signaling in terms of a modulation of synaptic drive, either at excitatory \citep{Savin_JRSI2009,Savtchenko_PTRSB2014} or excitatory and inhibitory synapses \citep{Ullah_JCN2009,Volman_PCB2013,Garnier_JMN2016}, which accounts for the emergence of variegated network activity. With this regard, a model by \citet{Ullah_JCN2009} considers the effect of astrocytic regulation of extracellular \K\ and \Na\ on firing dynamics of both excitatory and inhibitory neurons with voltage-dependent synaptic inputs. Accordingly, neuronal membrane potential dynamics depends on \K\ and \Na\ homeostasis regulated by astrocytes, and so does the network's E--I balance. In this framework, brief (\SI{30}{\milli \second}-long) step increases of E-to-E synaptic strength, which could loosely ensue from glial glutamate exocytosis, promote transient increases of neuronal firing, whose duration, however, depends on extracellular \K\ concentration \citep{Ullah_JCN2009}. In particular, for plausible non-pathological values of this concentration, the network could display transient episodes of persistent enhanced firing, which are reminiscent of persistent activity during UP~states \citep{SanchezVives_NN2000,Amzica_JN2002,McCormick_CC2003}, as well as during delay periods of working memory tasks \citep{Funahashi_JNP1989,GoldmanRakic_Neuron1995}.

Similar observations may also be made by other models which consider different scenarios of gliotransmission, such as short-term modulations of E-to-I synaptic connections by glutamatergic or purinergic gliotransmission \citep{Savtchenko_PTRSB2014}, or homeostatic upregulation of excitation by glial \tnfa, although this latter scenario could also account for the emergence of paroxysmal activity in various pathological conditions \citep{Volman_PCB2013}. Significantly, these models identify in the spatial extent of gliotransmission a further key factor for the regulation of glia-mediated episodes of increased network activity, ranging from their frequency and duration \citep{Volman_PCB2013} to the degree of network synchronization and the average firing rate of single neurons during their occurrence \citep{Savtchenko_PTRSB2014}. With this regard, the transient depression of synapses within an astrocytic anatomical domain, which could mimic release-decreasing gliotransmission or temporary disruption of release-increasing gliotransmission, correlates with a decrease of neuronal firing and synchronization, which is more significant for larger astrocytic domains \citep{Savtchenko_PTRSB2014}. 

\subsection{Regulation of synaptic transmission and plasticity}
Modulation of synaptic release probability by gliotransmitters may occur on multiple time scales  \citep{DePitta_Neurosci2015}. On the one hand, it may last for tens of seconds up to a few minutes, thus affecting synaptic transmission and network computations only temporarily. In this context, theoretical investigations hint that synapses that display short-term depression can turn facilitating, and vice verse, by gliotransmission \citep{DePitta_PCB11}. At the same time, the filtering characteristic of the synapse for incoming~APs also changes based on the history of astrocytic activation \citep{DePitta_Springer2019_Ch10}. On the other hand, it is also possible for gliotransmission and its effects on synaptic transmission to last for tens of minutes: that is on time scales that could promote long-term plastic changes of the synapse. In this latter scenario, the change of synaptic weight $w_{ij}$ of a synapse from neuron $j$ to neuron $i$ respectively firing at $x_j,\,x_i$, and in the presence of gliotransmitter release at rate $\gamma$, may be expressed by $\dot{w}_{ij} = F(w_{ij};x_i,x_j,\gamma)$, where~$F$ is a generic function which we can expand about $x_i=x_j=\gamma=0$, so that 
\begin{align}
\dot{w}_{ij} &\approx c_2^\mathrm{corr}x_i x_j + c_2^\mathrm{pre}x_i^2 + c_2^\mathrm{post}x_j^2\nonumber\\
             &\phantom{\approx}+c_2^\mathrm{g-pre}x_i\gamma + c_2^\mathrm{g-post}x_j\gamma + c_2^\mathrm{glia}\gamma^2 \nonumber\\
             &\phantom{\approx}+c_1^\mathrm{pre}x_i + c_1^\mathrm{post}x_j + c_1^\mathrm{glia}\gamma + c_0(w_{ij}) + \mathrm{O}(x_i,x_j,\gamma) \label{eq:learning-glia}
\end{align} 
\noindent
In the absence of gliotransmission ($c_2^\mathrm{g-pre}=c_2^\mathrm{g-post}=c_2^\mathrm{glia}=c_1^\mathrm{glia}=0$), the above equation results in prototypical \see{Hebbian Learning} by $c_2^\mathrm{corr}>0$ (or anti-Hebbian learning for $c_2^\mathrm{corr}<0$) \citep{Gerstner_BC2002}. This however may not be the case in the presence of gliotransmission, insofar as the requirements for associativity encompassed by Hebbian plasticity may not be sufficient to generate a change of synaptic weight. Correlation between pre- and postsynaptic activities reflected by $c_2^\mathrm{corr}$ in fact, now sums with correlations between these activities and the effect of gliotransmission on pre- ($c_2^\mathrm{g-pre}$) and postsynaptic receptors ($c_2^\mathrm{g-post}$) \citep{DePitta_NP2016}, including additional effects on synaptic weight borne by gliotransmission alone which, for example, could mirror SICs which could be independent of postsynaptic activity \citep{Wade_PLoSOne2011}.

Broadly speaking, how modulations of synaptic plasticity by glia could ultimately affect learning remains to be investigated, although few theoretical studies offer some enticing insights into the topic. Porto-Pazos and collaborators investigated performance of an astrocyte-inspired learning rule to train deep networks in data classification \citep{PortoPazos_PLoSOne2011,Alvarellos_CMMM2012,Mesejo_IJNS2015}. They consider a feedforward network architecture where each neuron in the intermediate (hidden) layers associate with an astrocyte that modulates all the neuron's outgoing connections. Each astrocyte is described by a quadruple $(\mu,\kappa,a,b)$ and requires $\mu$ stimulations by an associated synapse within $\kappa$ consecutive time steps to get activated. Upon an astrocyte's activation, the weights of all synapses associated with that astrocyte are increased by a factor~$a$ which could mimic either persistent increases of synaptic release by gliotransmission \citep{Perea_Science2007,Navarrete_PB2012} or LTP~following enhanced postsynaptic~NMDA receptor activation by astrocytically-released \dserine \citep{Henneberger_Nature2010}. Conversely, if a neuron associated with an astrocyte remains inactive $\mu$ times within $\kappa$ time steps, then the weights of its outgoing synapses are decreased by a factor~$b$ -- a behavior that loosely reflects the termination of the aforementioned gliotransmitter-mediated effects, and accounts, in the long run, for pruning of inactive synapses \citep{Hua_NN2004}.

Several variants of the learning rule were tested, mostly dealing with different handling of astrocyte activations by consecutive neuronal firing exceeding $\mu$; yet, regardless of the details of the learning (training) procedure, the trained neuron-glial networks were able to outperform identical networks without astrocytes in all discrimination tasks taken into account \citep{PortoPazos_PLoSOne2011,Alvarellos_CMMM2012}. The training was however successful if potentiation by astrocytes was less than depression by astrocyte inactivity, that is~$a<b$ \citep{Alvarellos_CMMM2012,Mesejo_IJNS2015}. Because the change in synaptic weight mediated by an astrocyte at a synapse~$i$ at time $t$ depends on the synapse's weight value at the previous update instant $t-1$, i.e. $\Delta w_i(t) = a\cdot w_i(t-1)$ for potentiation and $\Delta w_i(t) = b\cdot w_i(t-1)$ for depression, the $a<b$ condition assures that potentiation and depression do not cancel out at individual synapses, and that depression globally dominates, possibly preventing the network's runaway by excitation. On the other hand, the fact that $\Delta w_i(t)$ depends on $w_i(t-1)$, which in turn depends on $w_i(t-2)$ by $\Delta w_i(t-1)$ and so on, suggests that the relative contribution of potentiation over depression at any time at a given synapse depends on the history of synaptic updates which, in turn, reflects the history of the associated astrocyte's activation. 

There is circumstantial evidence that astrocytes could modify the threshold for~LTD vs.~LTP induction, such as in the supraoptic nucleus -- where astrocytic coverage of synapse is reduced during lactation \citep{Panatier_Cell2006}. In these conditions, stimuli that are expected to induce~LTP, elicit~LTD instead, possibly for a reduced~NMDA receptor activation by lower extracellular concentrations of astrocytic \dserine\ due to the increased extracellular space. Based on this experimental finding, and the observation that astrocytic \dserine\ could gate~LTD or~LTP induction \citep{Zhang_CC2008,Henneberger_Nature2010}, \citet{Philips_PCB2017} devised a modified version of the~BCM rule \citep{Bienenstock_JN1982,Gerstner_BC2002} where the threshold rate of postsynaptic firing for induction of~LTD vs.~LTP varies proportionally with astrocyte activation, and investigated how this rule affects development of orientation preference maps~(OPMs) in a self-organizing network model of the primary visual cortex~(V1) \citep{Stevens_JN2013}. In their formulation, activation of an astrocyte is computed by the weighted sum of activities of all synapses within the astrocyte's anatomical domain of radius~$R$, while the modified~BCM rule only applies to excitatory connections, whereas inhibitory connections and excitatory afferents are trained by a classical Hebbian paradigm. This allows choosing the spatial range of inhibitory connections to match short-distance lateral inhibitory connections found in~V1 \citep{Kisvarday_CC1997}, while leaving the spatial range of lateral excitatory connections dependent on astrocytic radii. 

This choice not only allows reproducing map orientation experimentally observed in~V1, but also reveals that, upon reduction of astrocytic radius, periodicity of~OPMs increases while width of individual hypercolumns decreases \citep{Philips_PCB2017}. Inasmuch as astrocyte size varies across species \citep{Oberheim_JN2009,Lopez_JCN2016}, these results predict a causal link between astrocytic radius and different hypercolumn widths observed in different species \citep{Kaschube_Science2010}. On the other hand, the model fails to explain why~V1 in rodents displays a typical salt-and-pepper configuration \citep{Ohki_Nat2005}, inasmuch as it predicts the emergence of such configuration in the limit of $R\rightarrow 0$, that is in the absence of astrocytes, which is in apparent contradiction with experimental evidence \citep{Schummers_STKE2008,Chen_PNAS2012,Perea_NC2014}.

The type of sliding threshold introduced by Philips and co-workers is conceptually different from the one in the original~BCM rule \citep{Bienenstock_JN1982}. In this latter, the sliding threshold is a local quantity that depends on the history of the activation of individual postsynaptic neurons. Conversely, by exploiting the concept of ``synaptic islands'' -- namely sets of synapses stimulating and being regulated by the same astrocyte \citep{Halassa_JN2007} -- Philips and colleagues \textit{de~facto} consider a ``global'' threshold that modulates weight of many synapses in parallel, based on the integral of their activation by their associated astrocyte. In this way, it is as if each astrocyte in \citeauthor{Philips_PCB2017}'s model supervised the signal inducing plastic changes (i.e., the ``teaching signal'') throughout the whole synaptic territory defined by its anatomical radius. 

The emerging view on \ca\ compartmentation in astrocytes \citep{Bazargani_NN2016} suggests that, if \ca\ is the mediator of gliotransmission \citep{Sahlender_PTRSB2014}, then its ``globality'' could be non-trivially defined, as it would likely depend, from time to time, on dynamics of \ca\ microdomains, rather than be solely defined by the astrocyte radius \citep{Volterra_NRN2014}. On the other hand, insofar as the criteria for globality are met, we shall note that the underpinning \ca\ signal could be generated either by postsynaptically released endocannabinoids \citep{Navarrete_PTRSB2014} or by different presynaptic pathways that are related to presynaptic neuronal firing. These presynaptic pathways could include the spillover of synaptically-released neurotransmitters from the synaptic cleft \citep{Araque_Neuron2014}, as well as other \ca\ pathways linked with extracellular ion homeostasis (\figref{fig:interactions}). Remarkably, if we consider this second option -- namely that the global \ca-dependent, gliotransmitter-mediated teaching signal carries information about the activity of presynaptic neurons --, it is possible to envisage a biologically plausible learning rule to learn precise spike patterns and reproduce them precisely and robustly over trials. This argument was elegantly demonstrated by \citet*{Brea_JN2013} in recurrent networks where neurons were arbitrarily separated between visible vs. hidden ones. In those networks, teaching to and accurate recall from visible neurons of an arbitrary spiking sequence is possible devising a learning rule that minimizes the Kullback-Leibler divergence of the spiking distribution produced by the network from the target distribution (i.e., the sequence to be learned). Such learning rule also discriminates between visible and hidden neurons and specifically modulates those synapses onto hidden neurons based on a threshold for LTD vs. LTP that is updated at each time step to account for the global activity of visible neurons. As the same authors argue that astrocytes could mediate this threshold, they also note that for this possibility to be realistic, astrocytes ``need to know'' which neurons are visible and which are hidden -- a scenario that could be brought forth by a combination of the reciprocal arrangement between astrocytes and neuronal and vascular structures, and the chemical signals between them yet to be identified.

\subsection{Glial cytokine signaling}
There is emerging evidence that, apart from macroglia, microglia, too, could be involved in the genesis and function of neural circuits in the healthy brain \citep{Kettenmann_Neuron2013}. This possibility is further supported by the growing recognition that molecules like proinflammatory cytokines such as \tnfa, which are generally associated with the microglial immunocompetent response, but could also be released from astrocytes \citep{Bezzi_NN2001}, may be found in healthy, non-inflamed brain tissue, possibly with a signaling role other than proinflammatory \citep{Wu_TiI2015}. Significantly, microglial could control extracellular \tnfa\ in a bimodal fashion, first increasing it up to a peak concentration, and then recovering its constitutive extracellular concentration \citep{Chao_DN1995}. Because at constitutive concentrations of $<$\SI{300}{\pico\Molar}, \tnfa\ seems necessary for glutamatergic gliotransmission, but above those concentrations, it could promote neurotoxicity \citep{Santello_TiNS2012}, modeling of \see{Cytokine Networks, Microglia and Inflammation} may be used to identify critical macro- and microglial pathways underpinning this dual signaling role by \tnfa. In this context, variance-based global sensitivity analysis of a microglial cytokine signaling network identifies two further cytokines -- \il10\ and \tgfb\ -- as possible key regulators of microglial \tnfa. Both those molecules are promoted by \tnfa\ while exerting negative feedback on it. However, because the kinetics of \il10-mediated feedback is faster than that by \tgfb, they can end up exerting opposing effects on extracellular \tnfa, depending on temporal differences in their expression: reduction of \tnfa\ peak concentration by \il10\ may indeed be counteracted by the ensuing reduction of \tgfb\ production with the potential to turn the traditionally anti-inflammatory action of these cytokines, proinflammatory instead \citep{Anderson_MBS2015}.

The study of \tnfa\ signaling, possibly of (micro)glial origin, is also linked with homeostatic mechanisms of synaptic plasticity \citep{Steinmetz_JN2010}. A case study is the mechanism of ocular dominance plasticity~(ODP) whereby monocular deprivation~(MD), during a critical period of development, causes a rearrangement of neuronal firing properties in the binocular visual cortex. In this fashion, cells that are initially biased to respond to inputs from the closed eye, end up responding more strongly to inputs from the open eye \citep{Wiesel_Nature1982}. As revealed by experiments in the visual cortex of juvenile mice, MD-triggered ODP results from two separable plastic processes: (i)~a rapid, Hebbian-like LTD, that is responsible for weakening the closed eye's response within the first three days of MD; and (ii)~a slow homeostatic response where \tnfa\ -- possibly of glial origin -- scales up excitatory synapses and strengthens the open eye's response approximately by the third day of MD \citep{Kaneko_Neuron2008}. Because of the large separation between time scales of Hebbian vs. homeostatic plasticity, conventional models of synaptic plasticity where Hebbian and homeostatic plasticity are assumed to compete to set synaptic strength, cannot account for MD-induced~ODP. In those models, in fact, a small perturbation of synaptic strength away from equilibrium is promptly amplified by the fast positive feedback of Hebbian plasticity, and this amplification can hardly be prevented by the slow homeostatic feedback, making the network prone to instability \citep{Toyoizumi_Neuron2014,Zenke_CON2017}. To avoid this scenario, an adequate description of MD-induced ODP should leverage instead of the consideration that observed Hebbian LTD and glial homeostatic scaling are independent of each other during MD \citep{Kaneko_Neuron2008} so that they both must separately reach their own stable state for the network to be stable. This could, in principle, be accounted for by the fact that plastic changes induced both by Hebbian learning and by homeostatic scaling can saturate to minimal and maximal values, making these two plasticity mechanisms inherently stable \citep{Beattie_Science2002,Connor_PNAS2005}. However, in practice, at least two further conditions must be met to be able to reproduce ODP robustly \citep{Toyoizumi_Neuron2014}. First, homeostatic plasticity must also scale minimum and maximum values of synaptic strength attained by mere Hebbian learning. Second, homeostatic scaling must depend on instantaneous neuronal activity, like postsynaptic depolarization, since this latter correlates with extracellular glutamate levels, which, in turn, regulate the glial release of \tnfa\ \citep{Bezzi_NN2001,Stellwagen_Nature2006}. Based on these arguments then, it is possible to devise a working model of MD-induced plasticity where synaptic strength $w$ can be recast as the product of two factors: a synapse-nonspecific factor $H$, applicable to the whole postsynaptic cell and controlled by slow glial \tnfa-mediated homeostatic scaling, and a synapse-specific factor $\rho$ regulated by fast Hebbian plasticity, i.e., $w=H\cdot\rho$ \citep{Toyoizumi_Neuron2014}. This description of synaptic strength then accounts for the intriguing possibility that slow glial \tnfa\ signaling in the visual cortex could intrinsically balance with fast Hebbian plasticity by instantaneously rescaling the range of synaptic strength values attained by Hebbian learning. The multiplicative scaling envisaged by the expression for $w$ is critical for glial homeostasis to maintain the relative strength between different synapses. Moreover, because it also affects minimal and maximal synaptic weights, it can thereby bring the network's activity level toward equilibrium without disturbing the intrinsic stability of the Hebbian dynamics and without being overwritten by this latter. In parallel, the presumed instantaneous dependence of $H$ on (post)synaptic activity guarantees network stability making glial homeostasis able to readily follow perturbations of the network's activity caused by fast Hebbian learning \citep{Toyoizumi_Neuron2014}.

\section{Temporal and spatial scales of neuron-glial interactions}
\subsection{Morphology and structural plasticity}
The large heterogeneity of glial cells, even within individual types such as oligodendrocytes, microglia, and astrocytes, makes the notion of average cell anatomy of little significance. The number of axons myelinated by single oligodendrocytes, for example, considerably changes with axon caliber, and so does the degree of myelination in terms of internodal lengths, ultimately dictating different conduction speeds. Oligodendrocytes are generally classified into four types depending on the caliber of the axons they myelinate, with the possibility for intermediate types to exist too. Accordingly, types~I/II are found in association with small ($<$\SI{4}{\micro \meter}) axons, whereas types~III/IV myelinate larger axons. The morphological values provided in \tabref{tab:scales} are from \citet{Butt_2013Ch}.

More complicated is the description of microglia since, in the healthy brain, these cells display a ramified morphology whose processes continuously extend and retract at rates of 0.4--\SI{3.8}{\micro\meter/\minute} \citep{Nimmerjahn_Science2005}. The classical study by \citet{Lawson_Neurosci1990} arguably still provides the most comprehensive characterization of microglia anatomy and distribution in the brain uptodate. Accordingly, microglia were shown to be present in large numbers in all major divisions of the brain but not to be uniformly distributed. There is, in fact, a more than five-fold variation in the density of microglial processes between different regions, with higher density in the hippocampus, average one in the cortex and hypothalamus, and lower one in the cerebellum. Overall, microglia surface density ranges between $\sim 50-\SI{140}{cells/\milli \meter^2}$, with individual cells displaying a large variability in surface area, from $<$\SI{200}{\micro \meter^2} in the cortex to $>$\SI{500}{\micro\meter^2} in the hippocampal formation, which likely associates with variegated morphological complexity, as reflected by a perimeter-to-convex perimeter ratio of~5 for cortical microglia but $>$8 for microglia in the hippocampus. Volumetric data on microglia is not yet available.

Protoplasmic astrocytes represent the main astrocyte type in the gray matter and appear to be among the most structurally intricate cells of the brain \citep{Reichenbach_2013}. They display a highly branched morphology with some degree of polarization that is probably region-specific \citep{Chao_Book2002}. At least one of the cell branches (or ``processes'') is bearing one or several perivascular endfeet such that the surfaces of the blood vessels in the~CNS are virtually completely ensheathed by astrocytic endfoot plates \citep{Mathiisen_Glia2010}. Astrocyte volume in rodents has been reported to range from~14700--\SI{22906}{\micro\meter^{3}} in the cortex to 65900--\SI{85300}{\micro\meter^{3}} in the hippocampus \citep{Bushong_JN2002,Ogata_Neurosci2002,Chvatal_JA2007,Halassa_JN2007}. These figures, however, are not indicative of the morphological complexity of these cells, as mirrored instead by a large surface-area-to-volume ratio in the range of 18.9--\SI{33.0}{\micro\meter^{-1}} \citep{Hama_JNC2004}. Single anatomical domains of cortical astrocytes were reported to include from~4 to~8 neuronal somata \citep{Halassa_JN2007}. A direct measure of the number of synapses within an astrocytic domain instead is not available but can be estimated from synaptic densities. With this regard considering an average of \SI{0.89}{synapse/\micro\meter^3} in the cortex \citep{Kasthuri_Cell2015} and \SI{2.13}{synapse/\micro\meter^3} in the hippocampus \citep{Kirov_JN1999}, we get figures for the number of synapses within the above astrocytic volumes that equal to $\sim$13000--20400 synapses/astrocyte in the cortex and $\sim$140000--182000 synapses/astrocyte in the hippocampus.

It is not clear to what extent the number of synapses ensheathed by astrocytes could change during activity by structural plasticity of perisynaptic astrocytic processes \citep{Haber_JN2006}, and cell swelling \citep{Florence_POne2012}, but, likely, the functional interactions between astrocytic and synaptic elements do so. Typical rate values for changes of cell morphology and ECS shrinkage ensuing from astrocytic swelling can then be estimated by experimental observations that directly link these changes with (putative) functionally relevant levels of neuronal activity. With this regard, perisynaptic astrocytic processes undergo multiple retractions and protrusions of $>$\SI{5}{\micro \meter} length in correlation with neuronal activation \citep{Haber_JN2006}, and~LTP induction correlates with an average remodelling rate of $<$\SI{40}{\nano \meter/\second} in hippocampal astrocytic processes \citep{Perez_JN2014}. Putative physiological \HB\ increases in the ECS trigger astrocytic swelling, which can be quantified in terms of percentage variation of the cell's surface area to baseline \HB\ concentrations. This swelling can be well described by a monoexponential curve of the type $\Delta_{\mathrm{max}}(1-\exp(-rt))$ with $r=\SI{0.08}{\second},\,\Delta_{\mathrm{max}}\approx 20\%$, resulting in an initial linear rate of surface area increase of $\sim$1.5~$\Delta$\%/\si{\second} ($\Delta$\%/\si{\second}: percentage variation in the unit time) \citep{Florence_POne2012}. Alternatively, ECS shrinkage possibly related to astrocyte swelling appears to increase linearly with duration of stimulation by rates in between $\sim$0.3--0.7~$\Delta$\%/\si{\second}, and recover to original volumes upon cessation of stimulation with decay times $>$\SI{10}{\second} \citep{Larsen_Glia2014}.

\subsection{Calcium signaling}\label{sec:calcium-estimates}
Study of \ca\ signaling in microglia and oligodendrocytes is still in its infancy and the only estimates of rise ($\tau_r$), decay times ($\tau_d$), and full width at half maximum~(FWHM) of this signaling pathway may be derived from purinergically-evoked \ca\ elevations in microglia \textit{in vivo} \citep{Brawek_SR2017,Brawek_CC2017} and oligodendrocyte \textit{in situ} \citep{James_CC2001}. Time constant are estimated by fitting experimental \ca\ traces ($c(t)$) by a difference of two exponential, i.e. $c(t) = C\rpar{\exp(-t/\tau_d)-\exp(-t/\tau_r)}$ where $C$ is a normalization factor equal to $ C = \big(T^{\frac{\tau_\mathrm{x}}{\tau_d}} - T^{\frac{\tau_\mathrm{x}}{\tau_r}} \big)^{-1}$ with $T=\tau_r/\tau_d$ and $\tau_\mathrm{x} = \tau_r\tau_d/(\tau_d-\tau_r)$. Accordingly, for oligodendrocytes (based on $n=20$ \ca\ traces): $\tau_r = 6.5\pm\SI{1.2}{\second},\, \tau_d = 6.7\pm\SI{1.1}{\second}$ and FWHM~=~$14.2\pm\SI{2.6}{\second}$; and for microglia ($n=6$): $\tau_r = 0.8\pm\SI{0.3}{\second},\, \tau_d = 5.0\pm\SI{0.1}{\second}$ and FWHM~=~$3.9\pm\SI{1.3}{\second}$. 

The vast majority of observations on glial \ca\ signaling has been made so far on astrocytes, although several aspects of this signaling remain to be investigated and cannot be challenged by present techniques \citep{Rusakov_NRN2015}. For example, it is not yet possible to resolve \ca\ signals in fine perisynaptic astrocytic processes (e.g., lamellae or filopodia), although we are now in the conditions of resolving \ca\ dynamics in three-dimensional space \citep{Bindocci_Science2017}. Generally speaking, when considering astrocytic \ca\ signals \textit{in vivo}, different factors beyond development and technical approaches must be taken into account, such as whether these signals were recorded in awake or asleep/anesthetized animals, what brain area they occurred in, if they are spontaneous or evoked by stimulation, and in this latter case, what stimulus protocol was adopted \citep{Rusakov_NRN2015}. In what follows, we only consider estimates for protoplasmic astrocytes, which constitute the main astrocyte type of the rodent's grey matter. The reader should keep in mind that \ca\ dynamics in other astrocyte types as well as in M\"{u}ller cells and Bergmann glia could be different \citep{Matyash_BRR2010}. 

In astrocytes, spontaneous somatic/whole cell \ca\ signals are rare ($\nu\approx 5.8-\SI{9.2}{\milli\hertz}$), and characterized by a rise time in the range of $\tau_r\approx 2-\SI{20}{\second}$, a decay time $\tau_d \approx 3-\SI{25}{\second}$, and a full width at half maximum FWHM~$\approx$~5--\SI{160}{\second} \citep{HiraseBuzsaki2004,NimmerjahnHelmchen2004,Wang_NN2006,Bindocci_Science2017}. Recent estimates of these quantities in microdomains of astrocytic processes and endfeet provide: $\nu\approx 6-\SI{8}{\milli \hertz}$, $\tau_r <0.7-\SI{5}{\second}$, $\tau_d <6-\SI{10}{\second}$, FWHM~$\approx$~0.35--\SI{3}{\second} in processes, and $\nu\approx 29-\SI{40}{\milli \hertz}$, $\tau_r <3-\SI{18}{\second}$, $\tau_d <2-\SI{30}{\second}$, FWHM~$\approx$~0.75--\SI{12}{\second} at endfeet \citep{Bindocci_Science2017}. The onset delay of these events can be remarkably short and follow neuronal stimulation within $<$0.3--\SI{3}{\second} \citep{Winship_JN2007,Bindocci_Science2017,Stobart_Neuron2018}. Significantly, there appears to exist a clear distinction in distribution of values of $\nu,\, \tau_r$ and $\tau_d$ with respect to \ca\ events whose~FWHM is below vs. above $\sim$\SI{1.5}{\second} \citep{Bindocci_Science2017}. Furthermore size of \ca\ microdomains in processes appears slightly larger during stimulation with respect to resting state, i.e. $60.7\pm \SI{24.3}{\micro \meter^3}$ vs. $40.50 \pm \SI{3.11}{\micro \meter^3}$, whereas a large dynamical richness of \ca\ events is reported at the somatic level, with average signaling volumes of the order of $4470 \pm \SI{1051}{\micro \meter^3}$, with largest events being in the order of $6375 \pm \SI{1106}{\micro \meter^3}$ (estimated range of $\sim$890--\SI{10000}{\micro \meter^3}) \citep{Bindocci_Science2017}. Intracellular \ca\ propagation may be estimated in the range of $\sim$16$\,\pm \SI{8}{\micro \meter/\second}$ \citep[][Supplementary Figure~5]{DiCastro_NN2011}, whereas speed of intercellular \ca\ waves, measured mostly in brain slices so far, is in the range of $\sim$8--\SI{20}{\micro \meter/\second} \citep{ScemesGiaume2006,Oberheim_JN2009}, although a recent study \textit{in vivo} reported on large \ca\ waves that could recruit $>$80 astrocytes, and propagate as fast as $61 \pm \SI{22}{\micro \meter/\second}$ \citep{Kuga_etal_JN2011}.

\subsection{Glutamate-glutamine cycle and astrocyte-to-neuron lactate shuttle}
Both GGC and ANLS rely on glutamate (and GABA) uptake by astrocytic transporters. The transporter currents for these two neurotransmitters recorded in hippocampal astrocytes display a marked time scale separation. In the adult mice, the 20\%--80\% time rise of astrocytic glutamate uptake was estimated in the range of 0.92--\SI{1.65}{\milli \second} \citep{Diamond_JN2005} although some investigators suggest a longer 10\%--90\% rise time, up to $5.98\pm \SI{0.38}{\milli \second}$ could also be reached \citep{Kinney_JNP2002}. Decay times are instead reported in the range of 4.64--\SI{5.43}{\milli \second}. GABA~uptake is substantially slower with 10\%--90\% rise and decay time constants respectively comprised between $>$410.5--\SI{543.83}{\milli \second} and $>$0.93--\SI{4.34}{\second} \citep{Kinney_JNP2002}. A further limiting factor of GGC and ANLS is substrate availability, for which little is known \textit{in vivo} so far. Nuclear magnetic resonance data in anesthetized rats suggest a rate of glutamine synthesis by astrocytes between $\sim$0.13--\SI{0.21}{\micro \mole/\minute\cdot \gram}, which is nearly half of the estimated consumption rate of glutamate by Krebs cycle, thus hinting glutamine synthesis as a major metabolic pathway in the rat cortex \citep{Sibson_PNAS1997,Rothman_ARP2003}. For lactate production by individual astrocytes, astrocytic intracellular NADPH signaling in response to synaptically-activated metabotropic glutamate receptors (mGluRs) can be adopted as the readout of different scenarios of neuronal activity and brain states in conditions of simulated hyperemia (high $p\ce{O2}$) or vasoconstriction (low $p\ce{O2}$) \citep{Gordon_Nature2008}. This is because the enzyme that is responsible of lactate synthesis -- i.e., lactate hydrogenase -- is NADPH-dependent (\figref{fig:interactions}, \textit{purple pathway}) so that increases and decreases of NADPH following mGluR-mediated astrocytic activation can directly be linked to intracellular lactate metabolism by astrocytes, resulting in estimates for rise and decay time constants for this latter in between 4.1--\SI{5.2}{\second}, with lower values associated with high $p\ce{O2}$ and vasoconstriction \citep{Gordon_Nature2008}.

\subsection{Gliotransmission}
Estimates of time scales of modulation by gliotransmitters should be based on experimental data (mostly \textit{in situ}) that provide information on the time evolution of the effect of gliotransmission on synaptic transmission. \tabref{tab:scales} includes such estimates for short-  and long-term effects of release-increasing glutamatergic gliotransmission, and short-term, release-decreasing purinergic gliotransmission \citep{Perea_Science2007,Covelo_eLife2018}. A biexponential function (\secref{sec:calcium-estimates}) can be used to fit time series data of synaptic release probability ($p(t)$) following astrocytic \ca\ activation (set at $t=0$), thereby obtaining: for glutamatergic gliotransmission $\tau_r>0.05-\SI{0.2}{\second}, \, \tau_d=5.9-\SI{6.7}{\second}$ ($n=2$, short-term); $\tau_r=6.2-\SI{28.3}{\second}, \, \tau_d=18.7-\SI{95.7}{\second}$ ($n=3$, long-term); for purinergic gliotransmission ($n=2$): $\tau_r=1.6-\SI{8.2}{\second}, \, \tau_d=8.2-\SI{12.0}{\second}$. It is also possibly to estimate the rate of \dserine-dependent synaptic potentiation in terms of percentage variation of synaptic strength in the unit time, under the assumption of a linear increase of synaptic strength in the presence of activity-dependent \dserine\ release from astrocytes. The range of values for this postsynaptic pathway of gliotransmission provided in \tabref{tab:scales} was estimated from \citet{Yang_PNAS2003} and \citet{Takata_JN2011}.

Concerning regulation of AMPA and \gabaa\ receptors by glial \tnfa\ \citep{Stellwagen_Nature2006}, The average rate of change for expression of postsynaptic AMPA and \gabaa\ receptors by glial \tnfa\ \citep{Stellwagen_Nature2006} may be derived from the ratio between the total percentage change or receptor expression with respect to control ($\Delta\%$) over the duration of \tnfa\ stimulus ($T_\upalpha$). In this fashion, for hippocampal receptors: $157\% \pm 15\%$ increase of AMPA receptors, and $12\% \pm 4\%$ decrease of \gabaa\ receptors were reported for \tnfa\ applications of $T_\upalpha=15-\SI{25}{\minute}$ \citep{Stellwagen_JN2005}, whereas a decrease of $25\%-30\%$ of \gabaa\ receptors was measured for $T_\upalpha=\SI{45}{\minute}$ \citep{Pribiag_JN2013}. These data translate into rate values of $\sim$5--12~$\Delta$\%/\si{\minute} and $\sim-$0.3--0.7~$\Delta$\%/\si{\minute}, respectively for AMPA and \gabaa\ changes. One should however keep in mind that the effect of \tnfa\ is region and/or cell-specific, insofar as the AMPA-to-NMDA ratio was shown to increase from~1.75 to~3 in hippocampal pyramidal cells in the presence of \tnfa\ ($T_\upalpha=1-\SI{2}{\hour}$), but to decrease from~3.7 to 2.5 at medium spiny neurons in the striatum \citep{Lewitus_JN2014}.

\subsection{Ion homeostasis}
Extracellular ion homeostasis depends on the brain region and evolves with ongoing neuronal activity. At the rat's optic nerve, glia \K\ buffering displays a typical bi-exponential behavior, with a rise time that may slightly increase from $\sim$\SI{0.83}{\second} to $\sim$\SI{1.6}{\second} as a \SI{10}{\hertz} stimulus is delivered either for \SI{1}{\second} or \SI{10}{\second} \citep{Ransom_JP2000}. Buffering time by re-equilibration of extracellular \K\ concentration, mostly by inward-rectifying \K\ channels in combination with NKPs, is typically large, being comprised between $3.9\pm\SI{1}{\second}$ and $18\pm \SI{0.6}{\second}$ \citep{Ransom_JP2000,DAmbrosio_JNP2002,Chever_JN2010}. Significantly, this buffering time may considerably shorten in the presence of glial \ca\ signaling, and thereby modulations of glial~NKPs, with both rise and decay time of extracellular \K\ converging to $\sim$0.9--\SI{5.0}{\second} \citep{Wang_PNAS2012,Wang_STKE2012}. There is little information on the spatial extent of glia-mediated \K\ buffering \textit{in vivo}, although intracellular recordings in glia pairs in the suprasylvan gyrus suggest that it is probably fast, with propagation speeds $>$\SI{30}{\micro \meter/second} \citep{Amzica_JN2002}. Intriguingly, intracellular \Na\ propagation in hippocampal astrocytes (although demonstrated so far \textit{in situ}) could be at least two-fold faster, but decay to 2--\SI{10}{\micro\meter/\second} for intracellular glial sites $>$\SI{60}{\micro\meter} far from the stimulus locus \citep{Langer_Glia2012}.

\subsection{Regulation of the blood flow}
\ca-dependent regulation of blood flow by astrocytes is dependent on brain state, possibly with vasoconstriction observed in the presence of hyperemia, and vasodilation in association with hypoemia \citep{Gordon_Nature2008}. Onset of astrocyte-mediated dilations or constrictions in different brain areas, generally occur within $<$1--\SI{3}{\second} from onset of astrocytic \ca\ signaling \citep{Zonta_NN2003,MulliganMcVicar_Nat2004,Otsu_NN2015}. On the other hand, astrocytes seem to promote vasoconstriction to a larger magnitude then vasodilation, since reported percentage decreases of blood vessel diameters mediated by astrocytes could be $>$20\%, whereas observed increases are generally $<$10\% \citep{Zonta_NN2003,MulliganMcVicar_Nat2004,Gordon_Nature2008}. This is mirrored by linear rates of percentage vessel diameter variation that range between 1.6--1.7~$\Delta$\%/\si{\second} for vasoconstriction \citep{MulliganMcVicar_Nat2004} but only between $>$0.05--0.3~$\Delta$\%/\si{\second} for vasodilation \citep{Zonta_NN2003}. Significantly, these figures depend on the number of astrocytic endfeet simultaneously activated on the same blood vessel, since, for example, a single endfoot could trigger a $9.1\pm 0.7\%$ vasoconstriction, but multiple endfeet could account for a three-fold larger reduction of vessel diameter \citep{MulliganMcVicar_Nat2004}.

\section{Conclusion}
Modeling of neuron-glial interactions is an emerging field of Computational Neuroscience. The ubiquity of these interactions and the possibility that they may occur within the time and spatial scales that are usually ascribed to neuronal and synaptic function, calls for a revision of current neuron-based modeling paradigms to include potentially relevant effects mediated by glial cells. The modeling arguments discussed in this chapter go in such direction, further hinting a fundamental design of neuronal circuits where their structure and function are possibly intertwined with that of glia, thereby challenging the traditional Neuron Paradigm of the brain, in favor of an extended, more realistic Neuron-Glial one.

\section{Acknowledgments}
Writing of this chapter was made possible by the generous support of the Junior Leader Fellowship Program by ``la Caixa" Banking Foundation (LCF/BQ/PI18/11630006), as well as by the support of the Basque Government through the BERC~2018-2012 program, and by the Spanish Ministry of Science, Innovation and Universities:~BCAM~Severo Ochoa Accreditation~SEV-2017-0718.

\section{Cross-references}
\see{Brain Energy Metabolism}\\
\see{Calcium-Dependent Exocytosis, Biophysical Models}\\
\see{Cytokine Networks, Microglia and Inflammation}\\
\see{Computational Modeling of Intracellular Astrocyte \ca\ Signals}\\
\see{Goldman-Hodgkin-Katz Equation}\\
\see{Hebbian Learning}\\
\see{Nernst Equation}\\
\see{Neural Mass Action}\\
\see{Short Term Plasticity, Biophysical Models}

%%%--------------------------------------------------------------------------
%% Figures
%%%--------------------------------------------------------------------------
\newpage
\section*{Figure captions}
\textbf{\figref{fig:interactions}}. Common interactions between astrocytes and glutamatergic synapses. \textit{Yellow pathway}: astrocytic calcium signaling; \textit{red pathway}: gliotransmission (both glutamatergic and purinergic); \textit{green pathway}: cytokine (\tnfa) signaling; \textit{turquoise pathway}: glutamate-glutamine cycle; \textit{blue pathway}: \K\ buffering; purple pathway: astrocyte-to-neuron lactate shuttle; \textit{brown pathway}: pH~buffering; \textit{orange pathway}: glutathione metabolism; \textit{brown pathway}:~ascorbic acid exchange; \textit{magenta pathway}: vascular coupling. Modified from \citet{DePitta_Springer2019_Ch1}. ($\dashrightarrow$) $\rightarrow$: (indirectly) promoting pathway; $\dashv$: inhibiting pathway. 20-HETE:~2--hydroxy-eicosatetraenoic acid; AA:~arachidonic acid; Ado:~adenosine; AMPA:~$\upalpha$-amino-3-hydroxy-5-methyl-4-isoxazolepropionic acid; ADP~(ATP):~adenosine diphosphate (triphosphate); \AP:~aquaporin channel type~4; ApN:~ectoaminopeptidase~N; ASC:~ascorbic acid (reduced vitamin~C) Best-1:~bestrophin-1 ion channel; CA:~carbonic anhydrase; ClBX:~\Cl/\HB\ exchanger; ClC:~chloride channel; Cys:~cysteine; CysGly:~cysteinylglycine; DHA:~dehydroascorbic acid (oxidized vitamin~C); EAAT:~excitatory amino acid transporter; EETs:~epoxyeicosatrienoic acids; ER:~endoplasmic reticulum; GABA:~$\upgamma$-Aminobutyric acid; GAT:~GABA transporter; GDOR:~glutathione-dependent dehydroascorbate reductase; GGS/PTG:~glycogen synthase/UTP--glycogen--phosphate uridylyltransferase; GJC:~gap junction channel; Glu:~glutamate; glucose-6P:~glucose 6-phosphate; GLUT:~glucose transporter; Gly:~glycine; GPhos:~glycogen phosphorylase; GR:~glutathione reductase; GS:~glutamine synthetase; GSH:~glutathione (reduced form); GSSG:~glutathione (oxidized form); $\upgamma$GT:~$\upgamma$-glutamyl transpeptidase; \ip3:~inositol 1,4,5-trisphosphate; iRCs:~ionotropic receptor channels; Kir:~inwardly rectifying \K\ channel; Lac:~lactate; LDH:~lactate dehydrogenase; MCT:~monocarboxylate transporter; mGluR:~metabotropic glutamate receptor; NAD$^+$ (NADH):~oxidized (reduced) nicotinamide adenine dinucleotide; NADP$^+$ (NADPH):~oxidized (reduced) nicotinamide adenine dinucleotide phosphate; NBC:~\Na-coupled bicarbonate transporter; NCX:~\Na/\ca\ exchanger; \ce{NH3}:~ammonia; NHX:~\Na/\H\ exchanger; NKClC: \Na-\K-\Cl co-transporter; NKP:~\Na/\K-ATPase pump; NMDA:~\textit{N}-Methyl-\textsc{d}-aspartate; \ce{P2Y1}:~metabotropic purinergic receptor; PAG:~glutaminase; PGs:~prostaglandins; PLA$_2$:~phospholipase~A$_2$; PPCs:~purine-permeable ion channels; Pyr:~pyruvate; SAClC: swell-activated chloride channel; SERCA:~(sarco)endoplasmic reticulum \ca-ATPase; SNAT:~\Na-coupled neutral amino acid transporter; SVCT2:~\Na-dependent vitamin~C transporter~2; TACE:~\tnfa-converting enzyme; \tnfa~(TNFR):~tumor necrosis factor alpha (receptor); TRPA:~transient receptor potential channel; VGCC:~voltage-gated \ca\ channel; VGKC:~voltage-gated \K\ channel; VGNC:~voltage-gated \Na\ channel.\\[1ex]

\noindent
\textbf{\figref{fig:schemes}}. Feedback and feedforward pathways in neuron-glial interactions. \textbf{A}~Presynaptic pathway of gliotransmission stimulated by presynaptically-released neurotransmitters; \textbf{B}~postsynaptic pathway of gliotransmission stimulated by presynaptically-released neurotransmitters; \textbf{C}~presynaptic pathway of gliotransmission triggered by postsynaptic endocannabinoid release; \textbf{D}~postsynaptic pathway of gliotransmission mediated by postsynaptic endocannabinoid release; \textbf{E}~modulation of glutamatergic gliotransmission by \tnfa. \textbf{F}~\K\ buffering and regulation of neuronal activity; \textbf{G}~glutamate-glutamine cycle~(GCC); \textbf{H}~astrocyte-to-neuron lactate shuttle~(ANLS); \textbf{I}~extracellular pH~homeostasis; \textbf{J}~glutathione synthesis; \textbf{K}~neurovascular coupling; \textbf{L}~glia-mediated regulation of action potential conduction. \ce{[Ca^{2+}]_i}: intracellular (cytosolic) calcium concentration; \ce{[Cys]_i} (\ce{[Gln]_i}):~intracellular cysteine (glutamine) concentration; \ce{[ECB]_e} (\ce{[TNF\alpha]_e}): extracellular endocannabinoid (\tnfa) concentration; \ce{[K^+]_e} (\ce{[H^+]_e}):~extracellular \K\ (\H) concentration; \ce{[Nt]_e} (\ce{[Glu]_e}): extracellular neurotransmitter (glutamate) concentration; $v_m$:~membrane potential; AP:~action potential; LPS:~lipopolysaccharide; SDF1$\upalpha$:~stromal cell-derived factor~1~alpha.

\newpage
\section*{Figures}
%%
%% Figure 1
\newpage
\clearpage

%% Figure 2
\newpage
\clearpage

%%%%%%%%%%%%%%%%%%%%%%%%%%%%%%%%%%%%%%%%%%%%%%%%%%%%%%%%%%%%%%%%%%%%%%%%%%%%
% Tables
%%%%%%%%%%%%%%%%%%%%%%%%%%%%%%%%%%%%%%%%%%%%%%%%%%%%%%%%%%%%%%%%%%%%%%%%%%%%
\newpage
\section*{Tables}
\setlength\LTcapwidth{\textwidth} % default: 4in (rather less than \textwidth...)
\setlength\LTleft{0pt}            % default: \parindent
\setlength\LTright{0pt}           % default: \fill
\begin{longtable}{p{0.65\textwidth} r l}
\caption{Some relevant spatial and temporal scales of glial signaling in the murine brain.}\\[1ex]
    %% Declarations
    \hline
    Description             &Range       &Units\T\\
    \hline \endfirsthead

    \caption[]{\emph{continued}}\\
    \hline
    Description             &Range       &Units\T\\
    \hline \endhead
    \hline \multicolumn{3}{r}{\emph{continued on the next page}} \endfoot
    \hline \endlastfoot
    %% Effective Table
    \multicolumn{3}{c}{\textsl{Morphological features}}\T\\
    \hline				  	
	Oligodendrocytes (type I/II)\T\\
	\rowgroup{Associated fiber diameter}	       &$<$4		&\si{\micro \meter}\\	
	\rowgroup{Number of myelinated axons per cell} &10--50      &--\\
	\rowgroup{Internodal length}				   &$\sim$10--350       &\si{\micro \meter}\\
	\rowgroup{Associated axonal conduction speed}  &$<$20		&\si{\micro \meter/\second}\\
	Oligodendrocytes (type III/IV)\\	
	\rowgroup{Associated fiber diameter}	       &$>$4		&\si{\micro \meter}\\	
	\rowgroup{Number of myelinated axons per cell} &1--10       &--\\
	\rowgroup{Internodal length}				   &$\sim$400--1000     &\si{\micro \meter}\\
	\rowgroup{Associated axonal conduction speed}  &80--120	    &\si{\micro \meter/\second}\\	
	Microglia\\
	\rowgroup{Surface density}                     &50--140     &\si{cells/\milli\meter^2}\\
	\rowgroup{Surface area}                        &$>$200--500 &\si{\micro\meter^2}\\
	\rowgroup{Perimeter-to-convex perimeter ratio} &$>$5--8     &--\\
    Astrocytes (protoplasmic type)\\
	\rowgroup{Cell volume}			          &14700--85300     &\si{\micro \meter^3}\\
	\rowgroup{Surface area-to-Volume ratio}   &18.9--33		    &\si{\micro \meter^{-1}}\\
	\rowgroup{Neurons per cell}				  &4--8				&--\\
	\rowgroup{Synapses per cell}              &13000--182000	&--\\
    \hline
    \multicolumn{3}{c}{\textsl{Structural plasticity}}\T\\	    
    \hline
    Process motility\\
    \rowgroup{Astrocytes}    				   &$<$40		&\si{\nano \meter/\second}\\
	\rowgroup{Microglia} 					   &0.4--3.6    &\si{\micro \meter/\minute}\\
	Astrocytic swelling\\
    \rowgroup{\HB-mediated percentage cell surface area increase} &$<$1.5 &$\Delta$\%/\si{\second}\\
    \rowgroup{Activity-dependent associated ECS shrinkage}  &$<$0.3--0.7  &$\Delta$\%/\si{\second}\\
	\rowgroup{ECS volume recovery time}						&$>$10        &\si{\second}\\    
	\hline
    \multicolumn{3}{c}{\textsl{Non-astrocytic \ca\ signaling}}\T\\
    \hline
    Oligodendrocytes (soma)\T\\
    \rowgroup{Rise time}					  &5.2--7.7			&\si{\second}\\
    \rowgroup{Decay time}					  &5.5--7.8			&\si{\second}\\
    \rowgroup{FWHM}							  &11.6--16.8		&\si{\second}\\
    Microglia (soma)\\
    \rowgroup{Rise time}					  &0.5--1.1			&\si{\second}\\
    \rowgroup{Decay time}					  &4.9--5.1			&\si{\second}\\	
    \rowgroup{FWHM}							  &2.6--5.2			&\si{\second}\\
	\hline
    \multicolumn{3}{c}{\textsl{Astrocytic \ca\ signaling}}\T\\
    \hline
    Soma\T\\
    \rowgroup{Frequency}					  &5.8--9.2			&\si{\milli \hertz}\\
	\rowgroup{Rise time}				      &2--20			&\si{\second}\\
	\rowgroup{Decay time}					  &3--25            &\si{\second}\\
	\rowgroup{FWHM}                           &5--160		    &\si{\second}\\
	Processes\\
    \rowgroup{Frequency}					  &6--8				&\si{\milli \hertz}\\
	\rowgroup{Rise time}					  &$<$0.2--5		&\si{\second}\\
	\rowgroup{Decay time}                     &6--10            &\si{\second}\\
	\rowgroup{FWHM}                           &0.35--3          &\si{\second}\\
	Endfeet\\
    \rowgroup{Frequency}					  &29--40			&\si{\milli \hertz}\\
	\rowgroup{Rise time}					  &3--18            &\si{\second}\\
	\rowgroup{Decay time}					  &2--30			&\si{\second}\\
	\rowgroup{FWHM}							  &0.75--12         &\si{\second}\\
	Onset delay (from stimulation)			  &$<$0.3--3		&\si{\second}\\
	Microdomain volume\\
	\rowgroup{Soma}							  &$\sim$890--10000 &\si{\micro\meter^3}\\	
	\rowgroup{Processes}					  &$\sim$37--85     &\si{\micro\meter^3}\\
	Propagation\\
	\rowgroup{Intracellular speed}			  &$\sim$8--24      &\si{\micro\meter/\second}\\
	\rowgroup{Intercellular speed}			  &$\sim$8--80      &\si{\micro\meter/\second}\\
    \hline
    \multicolumn{3}{c}{\textsl{GGC and ANLS}}\T\\	    
    \hline    
    Glutamate uptake\T\\
    \rowgroup{Rise time}						&0.9--6		&\si{\milli\second}\\
    \rowgroup{Decay time}						&4.6--5.4   &\si{\milli\second}\\
    GABA uptake rate\\				
    \rowgroup{Rise time}						&410--544   &\si{\milli\second}\\
    \rowgroup{Decay time}						&0.9--4.3   &\si{\second}\\
    Glutamine synthesis 						&0.13--0.21 &\si{\micro\mole/\minute}\\
    Lactate metabolism (rise/decay time)		&4.1--5.2	&\si{\second}\\	
    \hline
    \multicolumn{3}{c}{\textsl{Gliotransmission}}\\	    
    \hline
    Presynaptic pathway (glutamate, release-increasing)\T\\
    \rowgroup{Rise time}						&$>$0.05--28.3	&\si{\second}\\
    \rowgroup{Decay time}						&5.9--95.7      &\si{\second}\\
	Presynaptic pathway (purines, release-decreasing)\\
    \rowgroup{Rise time}						&1.6--8.2       &\si{\second}\\
    \rowgroup{Decay time}						&8.2--12        &\si{\second}\\
    Postsynaptic pathway\\
    \rowgroup{LTP rate (\dserine)}							&0.3--4.0	 &$\Delta$\%/\si{\minute}\\
    \rowgroup{\tnfa-mediated AMPA receptor variation}    	&$\pm$5--12  &$\Delta$\%/\si{\minute}\\	
    \rowgroup{\tnfa-mediated \gabaa\ receptor variation} 	&$-$0.3--0.7 &$\Delta$\%/\si{\minute}\\
    \hline
    \multicolumn{3}{c}{\textsl{Ion homeostasis}}\\	    
    \hline
    \K\ buffering\\
    \rowgroup{Rise time}                		&0.8--5.1	&\si{\second}\\
    \rowgroup{Decay time}                		&3.9--18.6	&\si{\second}\\
    \rowgroup{Propagation}						&$>$30      &\si{\micro\meter/\second}\\
    \Na\ spatial buffering						&$>$2--80	&\si{\micro\meter/\second}\\	
    \hline
    \multicolumn{3}{c}{\textsl{Vascular coupling}}\\	    
    \hline
    Vasodilation\\
    \rowgroup{Max \% variation upon astrocytic \ca\ stimulation} &5--8   &$\Delta$\%\\
    \rowgroup{Percentage vessel diameter change rate} &$>$0.05--0.37&$\Delta$\%/\si{\minute}\\
    Vasoconstriction\\
    \rowgroup{Max \% variation upon astrocytic \ca\ stimulation} &9--27.5&$\Delta$\%\\
    \rowgroup{Percentage vessel diameter change rate} &1.6--1.7  &$\Delta$\%/\si{\second}\\
    Onset delay from astrocyte activation		&$<$1--3		 &\si{\second}
%\end{tabularx}
\label{tab:scales}
%\end{tabularx}
\end{longtable}
\clearpage
%%%%%%%%%%%%%%%%%%%%%%%%%%%%%%%%%%%%%%%%%%%%%%%%%%%%%%%%%%%%%%%%%%%%%%%%%%%%%
%% Bibliography
%%%%%%%%%%%%%%%%%%%%%%%%%%%%%%%%%%%%%%%%%%%%%%%%%%%%%%%%%%%%%%%%%%%%%%%%%%%%%
\newpage
\bibliography{./depitta_ecns.bib}

\end{document}